\documentclass[10pt]{iopart}
\usepackage{graphicx,amssymb,iopams,ifthen,rotating}
\begin{document}
%%%%%%%%%%%%%%%%%%%%%%%%%%%%%%%%%%%%%%%%%%%%%%%
\title{Coil-helix transition of polypeptide at water-lipid interface}  

% \author{G.P. Sharma$^{1}$, Y.K. Reshetnyak$^{1}$, O.A. Andreev$^{1}$, 
% M. Karbach$^{2}$, and G. M{\"{u}}ller$^{1}$}
\author{Ganga P. Sharma$^{1}$, Yana K. Reshetnyak$^{1}$, Oleg
A. Andreev$^{1}$, Michael Karbach$^{2}$, and Gerhard M{\"{u}}ller$^{1}$}

\address{$^{1}$
  Department of Physics,
  University of Rhode Island,
  Kingston RI 02881, USA}
\address{$^{1}$ Fachgruppe Physik, Bergische Universit{\"{a}}t Wuppertal,
  D-42097 Wuppertal, Germany}

%\email[Gerhard M{\"{u}}ller]{gmuller@uri.edu}

\pacs{61.30.Hn, 87.15.Aa, 87.15.He, 87.15.Cc}

%%%%%%%%%%%%%%%%%%%%%%%%%%%%%%%%%%%%%%%%%%%%%%%
\begin{abstract}
  We present the exact solution of a microscopic statistical mechanical model
  for the transformation of a long polypeptide between an unstructured coil
  conformation and an $\alpha$-helix conformation.  The polypeptide is assumed
  to be adsorbed to the interface between a polar and a non-polar environment
  such as realized by water and the lipid bilayer of a membrane.  The
  interfacial coil-helix transformation is the first stage in the folding
  process of helical membrane proteins.  Depending on the values of model
  parameters, the conformation changes as a crossover, a discontinuous
  transition, or a continuous transition with helicity in the role of order
  parameter.  Our model is constructed as a system of statistically
  interacting quasiparticles that are activated from the helix pseudo-vacuum.
  The particles represent links between adjacent residues in coil conformation
  that form a self-avoiding random walk in two dimensions.  Explicit results
  are presented for helicity, entropy, heat capacity, and the average numbers
  and sizes of both coil and helix segments.
\end{abstract}

\maketitle

%%%%%%%%%%%%%%%%%%%%%%%%%%%%%%%%%%%%%%%%%%%%%%%
%
\section{Introduction}\label{sec:intro}
%
%%%%%%%%%%%%%%%%%%%%%%%%%%%%%%%%%%%%%%%%%%%%%%%
The folding mechanisms of water-soluble proteins from primary to secondary and
higher-order structures has been thoroughly investigated over many years.  In
the study of protein translocation pathways into and across cell membranes,
which is a very important active area of current research, one important
problem requiring further elucidation is the coil-helix transition that
accompanies the insertion of a polypeptide into a lipid bilayer.
The theoretical modeling of this ubiquitous process in biological matter is
fairly complex due to the heterogeneous environment in which conformational
changes occur and the simultaneity or rapid succession of conformational
change and translocation.  Experimental studies are limited to the small
selection of polypeptides that are water soluble and undergo controllable
insertion/folding and exit/unfolding processes.

The folding of all helical membrane proteins/peptides, independent of the
insertion mechanism, is governed by the formation of helical segments in the
lipid bilayer environment.  This process is driven by hydrophobic interactions
and hydrogen bonding \cite{PE01, WW99, ECC+03, LW04}.  Its two main steps are
the transformation from coil to interfacial helix and the insertion of the
helix into the membrane with transmembrane orientation.
Variants of the \emph{pH Low Insertion Peptide} (pHLIP) family are water
soluble and prove to be well suited for the investigation of
membrane-associated folding and unfolding \cite{RSAE07, WMT+13}, reversibly
driven by changes in pH.  A drop in pH leads to the protonation of negatively
charged side chains, which enhances the hydrophobicity of the peptide and
initiates the aforementioned two-step process of folding and insertion.  A
subsequent rise in pH reverses the process: the peptide unfolds and exits.
Recent experimental studies have already established important thermodynamic
and kinetic parameters of the peptide-membrane interaction \cite{RAS+08,
  AKW+10, KWW+12}.

\begin{figure}[h]
  \begin{center}
   \includegraphics[width=85mm]{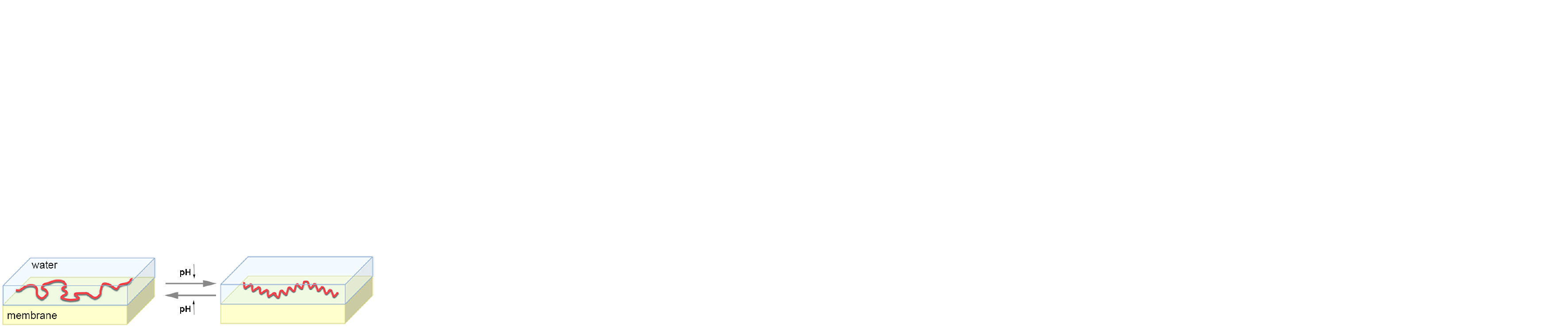}
\end{center}
\caption{(Color online) Long polypeptide at the interface between water and a
  flat lipid bilayer undergoing a reversible and pH-driven coil -helix
  transformation.}
  \label{fig:wat-lip-pep}
\end{figure}

What has been lacking for these and related experiments is a microscopic
statistical mechanical model with experimentally testable attributes that is
amenable to an exact analysis.  Our goal is to construct, solve, develop, and
test such a model in three successive stages.
The first stage, which is theme of this paper, involves the design and
solution of a microscopic model that describes the coil-helix transformation
of a long polypeptide adsorbed to the lipid bilayer of a membrane membrane
(see Fig.~\ref{fig:wat-lip-pep}).
%%%%%%%%%%%%%%%%%%%%%%%%%%%%%%%%%%%%%%%%%%%%%%%
%%%%%%%%%%%%%%%%%%%%%%%%%%%%%%%%%%%%%%%%%%%%%%%
Such a model is an indispensable part of a theory of membrane-associated
folding and will be used as the foundation for the next two stages.  They
include (a) the investigation of profiles of local attributes for generic
polypeptides and landscapes of global attributes for short peptides such as
pHLIP in the heterogeneous water/lipid environment and (b) the kinetics of
insertion and exit as can be inferred from the landscapes of free energy and
conformational attributes.

The pioneering theoretical studies of coil-helix transformations and related
phenomena that appeared throughout 1960s were admirably compiled and reviewed
in a monograph by Poland and Scheraga \cite{PS70}.  A series of model systems
were introduced at that time.  Many of them are still being used today in
textbooks and research papers.
This includes the familiar Zimm-Bragg model \cite{ZB59} and generalizations
thereof, all amenable to the transfer matrix method of analysis.  Also
emerging at that time was the highly original Lifson method of statistical
mechanical analysis \cite{Lif64}, the scope of which includes (smooth)
crossovers and (sharp) transitions \cite{PS66, Fish66}.  The need at this time
for yet another model solved by yet a different method is dictated by the
three stages of our project as will become apparent in what follows.

In Sec.~\ref{sec:micmod} we present the microscopic model of our design as a
system of statistically interacting links and describe the method of its exact
analysis.  Depending on the parameter settings the exact solution produces a
conformational change in the form of a crossover or a transition
(Sec.~\ref{sec:strusol}).  The transition may be of first or second order as
we discuss in Sec.~\ref{sec:ord-dis} with focus on the helicity (order
parameter) and entropy (measure of disorder) and other quantities.
In Sec.~\ref{sec:con-out} we summarize the main advances of this work and
point out their role as the foundation for the continuation of this project in
two different directions in the arenas of biological physics and statistical
mechanics.
We also discuss how this work connects to recent studies by other researchers and how the continuation of this project can benefit from those studies.

%%%%%%%%%%%%%%%%%%%%%%%%%%%%%%%%%%%%%%%%%%%%%%%
%
\section{Model system}\label{sec:micmod}
%
%%%%%%%%%%%%%%%%%%%%%%%%%%%%%%%%%%%%%%%%%%%%%%%
The microscopic model that we present here is a system of statistically
interacting quasiparticles with shapes.  The methodology employed for its
exact statistical mechanical analysis is built on the concept of fractional
statistics, invented by Haldane \cite{Hald91a}, and developed by Wu
\cite{Wu94}, Isakov \cite{Isak94}, Anghel \cite{Anghel}, and others
\cite{PMK07} in the context of quantum many-body systems.
The adaptation of this approach to classical statistical mechanical systems of
particles with shapes was developed in a recent series of studies with
applications to Ising spins \cite{LVP+08, copic, picnnn, pichs}, jammed
granular matter \cite{GKLM13,janac2}, lattice gases with long-range
interactions \cite{sivp}, and DNA under tension \cite{mcut1}.  The application
to the coil-helix transition of a long polypeptide adsorbed to a water-lipid
interface worked out in the following is conceptually simple but surprisingly
rich in scope.

%%%%%%%%%%%%%%%%%%%%%%%%%%%%%%%%%%%%%%%%%%%%%%%
\subsection{Coil segments from helix vacuum}\label{sec:he-co-seg}
%%%%%%%%%%%%%%%%%%%%%%%%%%%%%%%%%%%%%%%%%%%%%%%
The reference state (pseudo-vacuum) of our model system is the ordered helix
conformation of $N$ residues bound by peptide bonds into $N-1$ links and
stabilized by internal hydrogen bonds along the backbone.  Thermal
fluctuations or environmental change cause the nucleation of disordered coil
segments, which then grow by unravelling adjacent helical order.

In our model the coil segments are represented by thermally activated links
that combine to form a self-avoiding random walk between the ends of
successive helical segments.  Both coil and helix segments are confined to the
water-lipid interface (Fig.~\ref{fig:wat-lip-pep}).  Coil segments carry
configurational entropy that grows with their size and range in the interfacial plane. That range
is controllable at a microscopic level by the integer-valued model parameter $\mu$, henceforth 
called range parameter.

\begin{figure}[htb]
  \begin{center}
  \includegraphics[width=79mm]{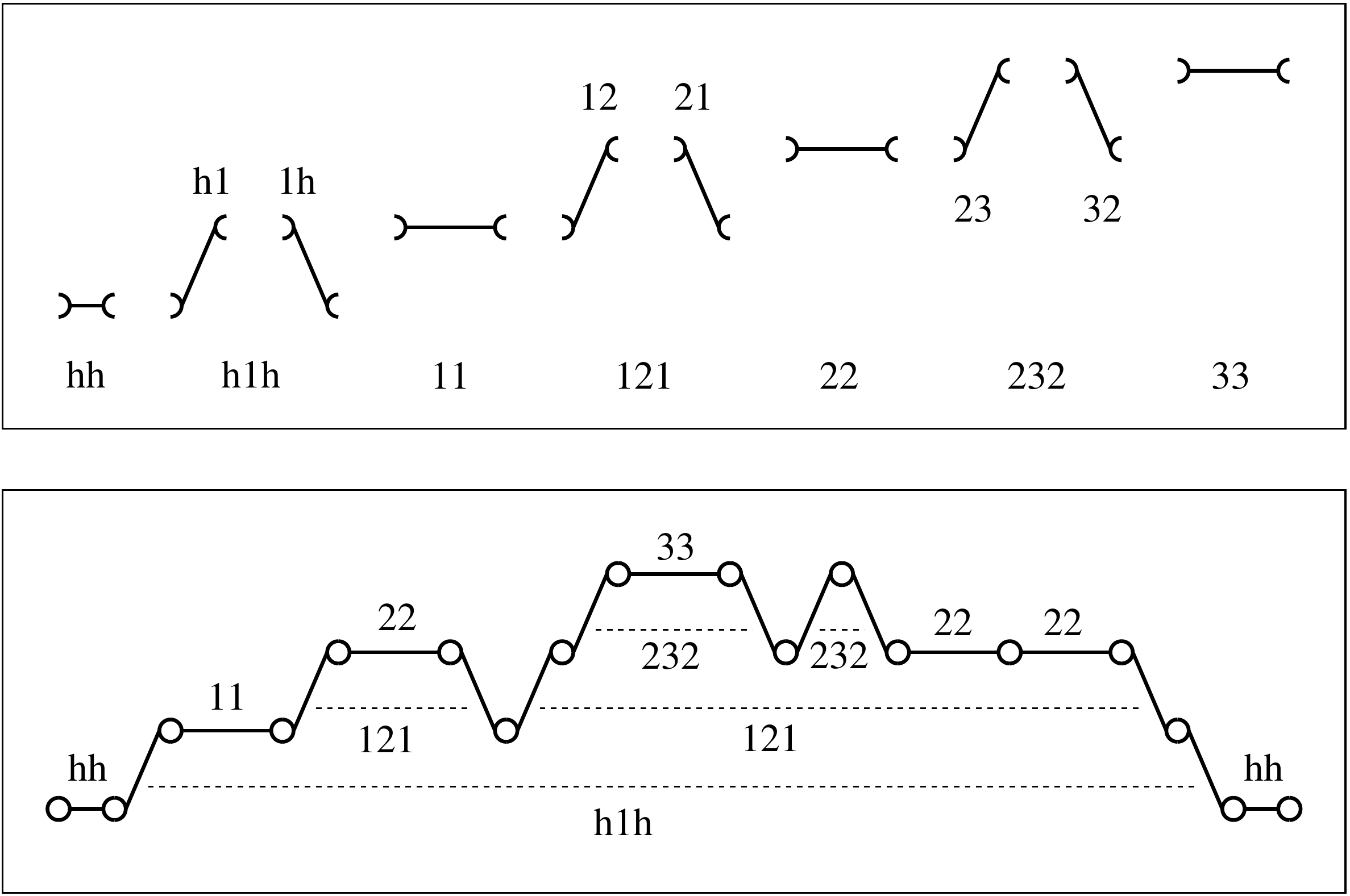}
\end{center}
\caption{Segment of coil conformation between two segments of helix
  conformation (bottom), generated by the activation of $2\mu$ species of
  statistically interacting particles in the form of single links or composed
  of pairs of links that are not necessarily adjacent (top).  Residues in
  helix conformation are in a unique state (h).  Residues in coil conformation
  are in one of $\mu$ states $(1, 2, . . . , \mu)$, constrained to to form a self-avoiding walk 
  in the interfacial plane and here illustrated for $\mu=3$.}
  \label{fig:links}
\end{figure}

Each residue can be in $\mu+1$ states, of which one (denoted \textsf{h}) represents the helix conformation and $\mu$ (numbered \textsf{1}, \ldots, $\mu$) represent the coil conformation.
Access by a residue to these states is constrained by the states of its
neighboring residues as illustrated in Fig.~\ref{fig:links} for the case
$\mu=3$.

%%%%%%%%%%%%%%%%%%%%%%%%%%%%%%%%%%%%%%%%%%%%%%%
%%%%%%%%%%%%%%%%%%%%%%%%%%%%%%%%%%%%%%%%%%%%%%%

Helical links (\textsf{hh}) are short and form straight segments (horizontal,
in Fig.~\ref{fig:links}) in the plane of the water-lipid interface.  Coil
links are more extended and directed either horizontally (\textsf{11, 22, 33})
or vertically (\textsf{h1, 12, 23, 32, 21, 1h}).  Our model allows each coil
segment to randomly explore the interface on one side of the helical direction
without intersecting itself.\footnote{The one-side restriction has no significant impact 
on the quantities we are calculating in this work.
Generalizations to coil segments that explore the plane of the interface more freely are planned.}
That space is discretized and
constrained by the length of the segment and by the number of distinct coil
states.

We assign different activation energies to coil links relative to \textsf{hh}
links depending on whether they contribute to nucleation or to growth.  To the
former we assign the activation energy $\epsilon_\mathrm{n}$ and to the latter
$\epsilon_\mathrm{g}$ (two model parameters).  Nucleation of one coil link
requires the simultaneous break-up of several internal hydrogen bonds whereas
growth proceeds by the break-up of one bond per link (two bonds shared by
different pairs of residues).  On our way to calculating a partition function
we now face the task of counting microstates of given link content.

%%%%%%%%%%%%%%%%%%%%%%%%%%%%%%%%%%%%%%%%%%%%%%%
\subsection{Combinatorics of links}\label{sec:comb}
%%%%%%%%%%%%%%%%%%%%%%%%%%%%%%%%%%%%%%%%%%%%%%%
For the combinatorial analysis we introduce a set of statistically interacting
quasiparticles that contain individual links or pairs of links.  It turns out
that we need $2\mu$ species of particles.  In the case of $\mu=3$ they are the
six species (along with the element of pseudo-vacuum) shown in the top panel
of Fig.~\ref{fig:links}.
The combinatorics of statistically interacting particles is captured by the
energy expression
\begin{equation}\label{eq:1} 
E(\{N_m\})=E_{\mathrm{pv}}+\sum_{m=1}^{2\mu} N_m\epsilon_m,
\end{equation}
and the multiplicity expression \cite{Hald91a, Wu94, Isak94, Anghel, PMK07,
  LVP+08,copic},
\begin{eqnarray}
  \label{eq:2a} 
  W(\{N_m\})=\prod_{m=1}^{2\mu}
  \left(\begin{array}{cc}
      d_m+N_m-1 \\ N_m    
    \end{array}\right),
  % \binom{d_m+N_m-1}{N_m},
  \\ \label{eq:2b} 
  d_m =A_m-\sum_{m'=1}^{2\mu} g_{mm'}(N_{m'}-\delta_{mm'}),
\end{eqnarray}
inferred from a generalized Pauli principle as will be illustrated below for
the application under consideration.  This means that there exist $W(\{N_m\})$
microstates with energy $E(\{N_m\})$ that all have the same particle content:
$N_m$ particles of species $m$ for $m=1,\ldots,2\mu$.  The $\epsilon_m$ are
particle activation energies relative to the energy $E_\mathrm{pv}$ of the
pseudo-vacuum, the $A_m$ are capacity constants, and the $g_{mm'}$ are
statistical interaction coefficients.  The specifications for the model with
$\mu=3$ distinct coil states are compiled in Table~\ref{tab:1}.

%%%%%%%%%%%%%%%%%%%%%%%%%%%BEGIN TABLE%%%%%%%%%%%%%%%
\begin{table}[htb]\small
  \caption{Specifications of the six species of particles that describe the case $\mu=3$.}\label{tab:1} 
   \begin{tabular}{ccc|cc}\hline\hline  \rule[-2mm]{0mm}{6mm}
motif & cat. & $m$ & $\epsilon_{m}$ & $A_{m}$ 
\\ \hline \rule[-2mm]{0mm}{6mm}
$\mathsf{h1h}$ & host & $1$ & $\epsilon_\mathrm{n}$ & $N-2$ 
\\ \rule[-2mm]{0mm}{5mm}
$\mathsf{121}$ & hybrid & $2$ & $2\epsilon_\mathrm{g}$ & $0$ 
\\ \rule[-2mm]{0mm}{5mm}
$\mathsf{232}$ & hybrid & $3$ & $2\epsilon_\mathrm{g}$ & $0$ 
\\ \rule[-2mm]{0mm}{5mm}
$\mathsf{11}$ & tag & $4$ & $\epsilon_\mathrm{g}$ & $0$ 
\\ \rule[-2mm]{0mm}{5mm}
$\mathsf{22}$ & tag & $5$ & $\epsilon_\mathrm{g}$ & $0$ 
\\ \rule[-2mm]{0mm}{5mm}
$\mathsf{33}$ & tag & $6$ & $\epsilon_\mathrm{g}$ & $0$ \\ \hline\hline 
\end{tabular}\hspace{2mm}
\begin{tabular}{c|rrrrrr}\hline\hline  \rule[-2mm]{0mm}{6mm}
$g_{mm'}$ & $1$ & $2$ & $3$ & $~~4$  & $~~5$ & $~~6$ \\ \hline \rule[-2mm]{0mm}{6mm}
$1$ & $2$ & $2$ & $2$ & $1$& $1$ & $1$\\ \rule[-2mm]{0mm}{5mm}
$2$ & $-1$ & $0$ & $0$ & $0$ & $0$ & $0$\\  \rule[-2mm]{0mm}{5mm}
$3$ & $0$ & $-1$ & $0$ & $0$ & $0$ & $0$\\ \rule[-2mm]{0mm}{5mm}
$4$ & $-1$ & $-1$ & $0$ & $0$ & $0$ & $0$\\  \rule[-2mm]{0mm}{5mm}
$5$ & $0$ & $-1$ & $-1$ & $0$ & $0$ & $0$\\ \rule[-2mm]{0mm}{5mm}
$6$ & $0$ & $0$ & $-1$ & $0$ & $0$ & $0$\\ \hline\hline 
\end{tabular} 
\end{table} 
%%%%%%%%%%%%%%%%%%%%%%%%%%%END TABLE%%%%%%%%%%%%%%

The particles form nested structures as indicated in Fig.~\ref{fig:links}.  We
have species from three categories in the taxonomy of Ref.~\cite{copic}: one
species of \emph{hosts}, two species of \emph{hybrids}, and three species of
\emph{tags}.  Hosts cannot be hosted, tags cannot host any particles from a
different species, hybrids can do both.

A system of $N$ residues in the helix pseudo-vacuum has the capacity of
nucleating a coil segment at $A_1=N-2$ different locations by activating a
host particle with activation energy $\epsilon_1=\epsilon_\mathrm{n}$.  The
activation of particles from any species reduces the capacity $d_1$ of the
system for placing further hosts on account of (\ref{eq:2b}) and $g_{1m}>0$.
Hosts and hybrids have twice the size of tags.  The former thus reduce the
capacity at double the rate of the latter.

For any mix of particles the system has a finite capacity.  When that capacity
has been reached, we have $d_1=1$, which makes the associated binomial factor
in (\ref{eq:2a}) equal to one.  If we attempt to add one more particle from
any species, $d_1$ becomes zero or a negative integer and, in consequence, the
associated binomial factor vanishes.
The helix pseudo-vacuum has zero capacity for the placement of hybrids and
tags, as implied by $A_2=\cdots =A_6=0$. Such capacity of the system is
generated dynamically by the placement of particles with hosting capacity.
This generation of capacity is encoded in the negative interaction
coefficients.  Hosts 1 generate capacity for placing hybrids 2 and tags 4.
Hybrids 2, in turn, generate capacity for placing hybrids 3 and tags, 4, 5
etc.  Tags do not generate capacity for placing any particles.

The particle content of the coil segment of $N=16$ residues shown in
Fig.~\ref{fig:links} is $N_1=1$, $N_2=2$, $N_3=2$, $N_4=1$, $N_5=3$, $N_6=1$.
Its activation energy (\ref{eq:1}) thus becomes $E(1,2,2,1,3,1)-E_\mathrm{pv}
=\epsilon_\mathrm{n} +13\epsilon_\mathrm{g}$ and the number of coil segments
of equal contour length and with the same particle content is, according to
(\ref{eq:2a}), (\ref{eq:2b}), ${W(1,2,2,1,3,1)=360}$.
Further microstates with equal activation energy are generated if we exchange
hybrids or tags from one species by hybrids or tags from a different species
or if we replace hybrids by pairs of tags (or vice versa), all within the
constraints imposed by the nesting.  The constraints are encoded in the
multiplicity expression.  It does not allow spurious particle combinations.

The generalization to any $\mu$ is straightforward: the zoo of $2\mu$ particle
species now comprises one host, $\mu-1$ hybrids, and $\mu$ tags, labeled
consecutively in this order.  The host has activation energy
$\epsilon_\mathrm{n}$, reflecting the nucleation of coil segments, whereas
hybrids and tags have activation energies $2\epsilon_\mathrm{g}$ and
$\epsilon_\mathrm{g}$, respectively, reflecting the growth of coil segments.
The capacity constants remain the same for each category:
$A_m=(N-2)\delta_{m,1}$.  

The nonzero interaction coefficients generalize
naturally in accordance with the sizes and nested structure of the particles:
$g_{1m}=2~(1)$ for $m=1,\ldots,\mu$ $(\mu+1,\ldots,2\mu)$; $g_{m'm}=-1$ for
three sets of index pairs: (i) $m=m'-1$, $m'=2,\ldots,\mu$; (ii) $m=m'-\mu$,
$m'=\mu+1,\ldots,2\mu$; (iii) $m=m'-\mu+1$, $m'=\mu+1,\ldots,2\mu-1$.
The case $\mu=1$ has no hybrids: $g_{11}=2$, $g_{12}=1$, $g_{21}=-1$, $g_{22}=0$.  It is equivalent to the Zimm-Bragg model
\cite{ZB59}.  With the combinatorial analysis completed we turn to the
statistical mechanical analysis.

%%%%%%%%%%%%%%%%%%%%%%%%%%%%%%%%%%%%%%%%%%%%%%%
\subsection{Free energy of polypeptide}\label{sec:stat-mech}
%%%%%%%%%%%%%%%%%%%%%%%%%%%%%%%%%%%%%%%%%%%%%%%
The partition function for the adsorbed polypeptide, modeled as a system of
statistically interacting and thermally activated particles \cite{Hald91a,
  Wu94, Isak94, Anghel, PMK07, LVP+08,copic},
\begin{equation}\label{eq:3} 
  Z=\sum_{\{N_m\}}W(\{N_m\})e^{-\beta E(\{N_m\})},
\end{equation}
depends on energy (\ref{eq:1}) and multiplicity (\ref{eq:2a}) with ingredients
$\epsilon_m, A_m, g_{mm'}$ from Sec.~\ref{sec:comb}.  The thermal equilibrium
macrostate in the thermodynamic limit follows from an extremum principle.  Its
implementation yields the partition function for a macroscopic system in the
(generic) form \cite{Wu94, Isak94, Anghel, PMK07, LVP+08, sivp},
\begin{equation}\label{eq:4} 
Z=\prod_{m=1}^{M}\big(1+w_m^{-1}\big)^{A_m},
\end{equation}
where $M=2\mu$ in our case and the (real, positive) $w_m$ are solutions of the
coupled nonlinear equations,
\begin{equation}\label{eq:5} 
e^{\beta\epsilon_m}=(1+w_m)\prod_{m'=1}^{M} \big(1+w_{m'}^{-1}\big)^{-g_{m'm}}.
\end{equation}
The average number of particles from species $m$ are derived from the coupled
linear equations,
\begin{equation}\label{eq:6} 
  w_m N_m+\sum_{m'=1}^{M}g_{mm'}N_{m'} =A_m.
\end{equation}

It is useful and economical to express all results as functions of the two
control parameters\footnote{The relevant energy scales are the strength of hydrogen bonds 
($\sim5$ kcal/mol) and $k_\mathrm{B}T$ at room temperature ($\sim0.6$ kcal/mol).}
% \cite{note2}
\begin{eqnarray}\label{eq:7} 
  & \tau\doteq e^{\beta(\epsilon_\mathrm{g}-\epsilon_\mathrm{n})} 
  & \quad : 0\leq\tau\leq1, \\
  \label{eq:8} 
  & t\doteq e^{\beta\epsilon_\mathrm{g}} 
  & \quad : 0\leq t<\infty,
\end{eqnarray}
with an additional dependence on the discrete range parameter $\mu$ implied.
The nucleation parameter $\tau$ is a measure of cooperativity and controls the
average length of coil and helix segments.  High cooperativity $(\tau\ll 1)$
means a high nucleation threshold.  Low cooperativity $(\tau \lesssim 1)$
means little difference in enthalpic cost of nucleation and growth.  The
growth parameter $t$ controls the preference of one or the other
conformation. Coil is preferred at small $t$ and helix at large $t$.

Equations (\ref{eq:5}) for $w_1,\ldots,w_{2\mu}$ with parameters $t,\tau$ used
on the left and the $g_{mm'}$ from Sec.~\ref{sec:comb} used on the right can
be reduced to a single polynomial equation of order $\mu+1$ for $w_{\mu+1}$:
\begin{equation}\label{eq:9} 
  (1+w_{\mu+1}-t)S_{\mu}(w_{\mu+1})=t\tau S_{\mu-1}(w_{\mu+1}),
\end{equation}
where the $S_m(w)$ are Chebyshev polynomials of the second kind.  Among all
the solutions of Eq.~(\ref{eq:9}) there exists exactly one,
\begin{equation}\label{eq:36} 
w\doteq w_{\mu+1}(t,\tau),
\end{equation}
that satisfies the criterion of physical relevance, requiring that
(\ref{eq:36}) and all the remaining $w_m$ inferred from it via
\begin{eqnarray}\label{eq:10} 
  & w_{1}=\frac{S_{\mu}(w)}{\tau S_{\mu-1}(w)}=\frac{t}{1+w-t}, 
  \nonumber \\
  & w_m=
  \left\{\begin{array}{cl}
  % \begin{cases}
    {\displaystyle \frac{S_{\mu-m+2}(w)}{S_{\mu-m}(w)}} & : m=2,\ldots,\mu, 
    \\
    w & : m=\mu+1,\ldots,2\mu,
%  \end{cases}
  \end{array}\right.
\end{eqnarray}
are non-negative.  The derivation of this reduction is outlined
in~\ref{sec:app-a}.

The Gibbs free energy per residue inferred from (\ref{eq:4}) then depends on
that physical solution as follows:
\begin{equation}\label{eq:11} 
\bar{G}(t,\tau) =-k_\mathrm{B}T\ln\big(1+w_{1}^{-1}\big), 
\end{equation}
from which any thermodynamic quantity of interest can be derived, including
the entropy,
\begin{eqnarray}\label{eq:12} 
  \bar{S} \doteq -\left(\frac{\partial\bar{G}}{\partial
      T}\right)_{\epsilon_\mathrm{n},\epsilon_\mathrm{g}},
\end{eqnarray}
the enthalpy,
\begin{equation}\label{eq:13} 
 \bar{H}
 \doteq \bar{G} +T\bar{S},
\end{equation}
the helicity (order parameter),
\begin{eqnarray}\label{eq:14} 
  \bar{N}_\mathrm{hl} \doteq 1-\left(\frac{\partial
      \bar{G}}{\partial\epsilon_{\mathrm{n}}}\right)_{T, \epsilon_\mathrm{g}}-
  \left(\frac{\partial \bar{G}}{\partial\epsilon_{\mathrm{g}}}\right)_{T,
    \epsilon_\mathrm{n}},
\end{eqnarray}
the density of (helix or coil) segments,
\begin{equation}\label{eq:15} 
  \bar{N}_\mathrm{seg}
  \doteq\left(
    \frac{\partial \bar{G}}{\partial\epsilon_{\mathrm{n}}}
  \right)_{T, \epsilon_\mathrm{g}},
\end{equation}
and the average sizes of helix segments and coil segments,
\begin{equation}\label{eq:16} 
  L_\mathrm{hs}\doteq \frac{\bar{N}_\mathrm{hl}}{\bar{N}_\mathrm{seg}},
  \quad 
  L_\mathrm{cs}\doteq \frac{1-\bar{N}_\mathrm{hl}}{\bar{N}_\mathrm{seg}}.
\end{equation}
The population densities $\bar{N}_m\doteq N_m/N$, $m=1,\ldots,2\mu$, of
particles can be extracted from the solution of the
linear Eqs.~(\ref{eq:6}) as shown in~\ref{sec:app-b}.

%%%%%%%%%%%%%%%%%%%%%%%%%%%%%%%%%%%%%%%%%%%%%%%
%
\section{Structure of solution}\label{sec:strusol}
%
%%%%%%%%%%%%%%%%%%%%%%%%%%%%%%%%%%%%%%%%%%%%%%%
Changing the level of pH primarily affects the growth parameter $t$.  At
normal pH we have $t\lesssim1$, which favors the random coil conformation.  A
drop in pH pushes the growth parameter to higher values, $t>1$, which
increasingly favors a conformation with helical ordering.\footnote{The level of pH effectively controls how easy or hard it is to replace broken internal hydrogen bonds along the backbone of the polypeptide with external hydrogen bonds involving $\mathrm{H_2O}$ molecules. 
Hence the shift in $t$.}
Depending on the value of the nucleation parameter $\tau$ and the discrete
parameter $\mu$, which controls the amount of entropy that coil segments can
generate, the growth of helicity takes place in a crossover or in a transition
of first or second order.  To illuminate the criteria for these alternatives
we investigate the nature of the physically relevant solution (\ref{eq:36}) of
Eq.~(\ref{eq:9}), in particular the singularities it acquires in the limits
$\tau\to0$ at $\mu<\infty$ and $\mu\to\infty$ at $\tau>0$.

%%%%%%%%%%%%%%%%%%%%%%%%%%%%%%%%%%%%%%%%%%%%%%%
\subsection{Crossover}\label{sec:cross}
\begin{figure}[t]
  \begin{center}
    \includegraphics[width=40mm]{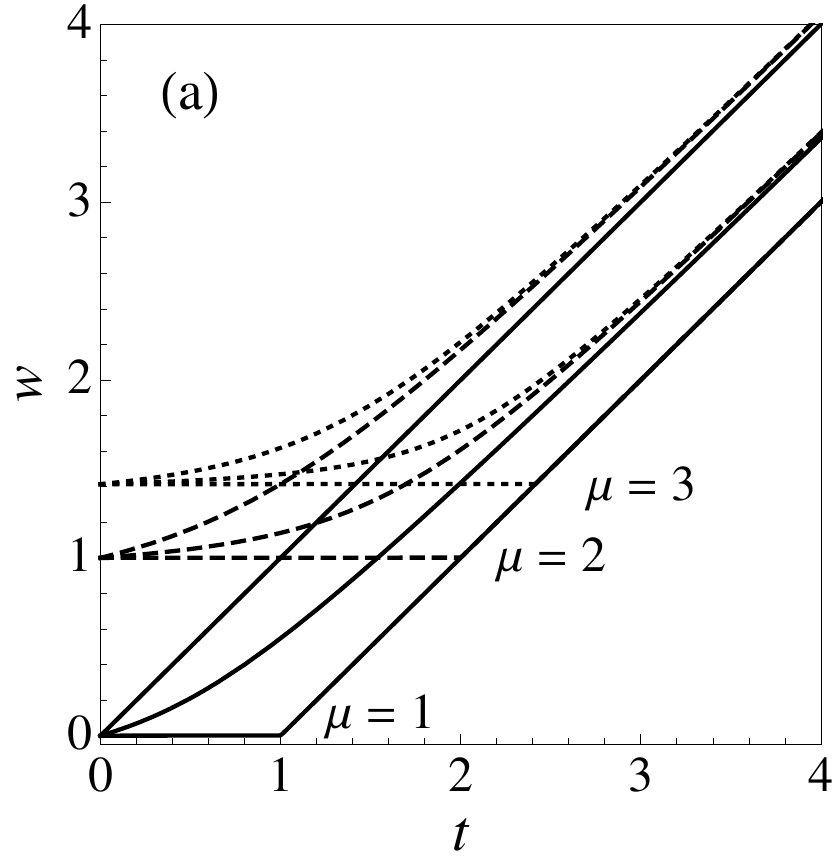}\hspace*{3mm}
    \includegraphics[width=40mm]{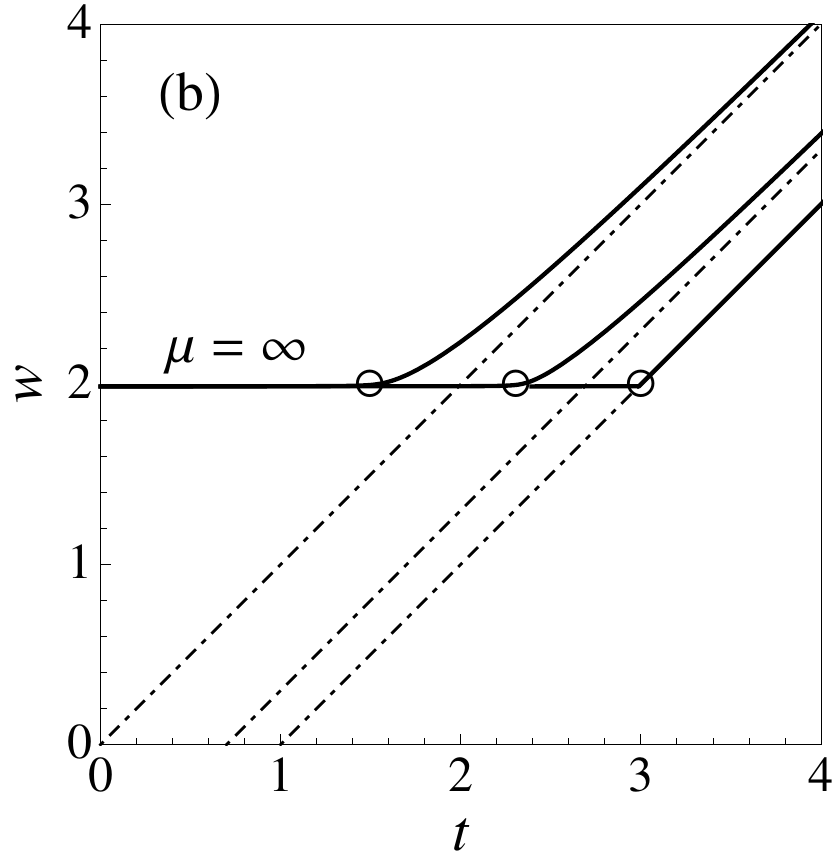}
\end{center}
\caption{Physical solution (\ref{eq:36}) of the polynomial equation
  (\ref{eq:9}) for $\tau=1, 0.3, 0$ (left to right in both panels).  Panel (a)
  shows the emergence of a singularity in the limit $\tau\to0$ for $\mu=1,2,3$
  and panel (b) the emergence of a singularity in the limit $\mu\to\infty$ at
  $\tau\geq0$.}
  \label{fig:crossover}
\end{figure}

%%%%%%%%%%%%%%%%%%%%%%%%%%%%%%%%%%%%%%%%%%%%%%%
For $\tau>0$ and $\mu<\infty$ the solution $w(t,\tau)$ is bounded from below
by
\begin{equation}\label{eq:18} 
w_0\doteq  2\cos\left(\frac{\pi}{\mu+1}\right),
\end{equation}
which is the location of the last zero of $S_\mu(w)$.  That value is only
realized at $t=0$ as illustrated in Fig.~\ref{fig:crossover}(a).  For $t\gg1$
the solution converges toward the linear asymptote,
\begin{equation}\label{eq:19} 
w_{\mathrm{as}}\doteq t+\tau-1.
\end{equation}
Note that $w_0$ depends on $\mu$ but not on $\tau$ whereas $w_\mathrm{as}$
depends on $\tau$ but not on $\mu$.  The smooth dependence on $t$ of
$w(t,\tau)$ for $\tau>0$ and $\mu<\infty$ describes a crossover from low
helicity at small $t$ to high helicity at large $t$.

%%%%%%%%%%%%%%%%%%%%%%%%%%%%%%%%%%%%%%%%%%%%%%%
%%%%%%%%%%%%%%%%%%%%%%%%%%%%%%%%%%%%%%%%%%%%%%%

The Zimm-Bragg model \cite{ZB59} is represented by the case $\mu=1$, for which
(\ref{eq:9}) is a quadratic equation with physical solution
\begin{equation}\label{eq:37}
w=\frac{1}{2}\left[t-1+\sqrt{(t-1)^2+4t\tau}\,\right].
\end{equation}
It is mathematically equivalent to an Ising chain.\footnote{The Zimm-Bragg parameters 
commonly used are 
$\sigma=\tau$ and $s=t$. The (physically relevant) larger eigenvalue of the transfer matrix is 
$\lambda_0=w+1$, taking into account a shift in energy scale by $\epsilon_\mathrm{g}$ per residue.}

The case $\mu=2$ is qualitatively different in that each coil segment now
carries entropy.  The physical solution of the associated (cubic)
Eq.~(\ref{eq:9}) reads
\begin{eqnarray}\label{eq:38} 
  &w=\frac{1}{3}\left[x+2\sqrt{x^2+3y}\cos\frac{\varphi}{3}\right], \\
  &\tan\varphi=\frac{\sqrt{27(4y^3+y^2x^2+18yx^2+4x^4-27x^2)}}
{x(2x^2+9y-27)},\nonumber
\end{eqnarray}
where $x\doteq t-1$, $y\doteq 1+t\tau$, and $0\leq\varphi<\pi$.

For $3\leq\mu<\infty$ and $\tau>0$ the solution (\ref{eq:36}) must be
determined numerically.  In that context it is advisable to rewrite
(\ref{eq:9}) as
\begin{equation}\label{eq:53} 
  (w+1-t)r_\mu(w)-t\tau=0,\quad r_\mu(w)\doteq\frac{S_\mu(w)}{S_{\mu-1}(w)},
\end{equation}
using the function
\begin{equation}\label{eq:54} 
  r_\mu(w)= \left\{ \begin{array}{ll}
      % \begin{cases}
      {\displaystyle \frac{1}{2}\left[w+\sqrt{4-w^2}\,\right]
        \cot\left(\mu\,\mathrm{arccos}\frac{w}{2}\right)} 
      & : w<2, \\
      {\displaystyle \frac{\mu+1}{\mu}} 
      & : w=2, \\
      {\displaystyle \frac{1}{2}\left[w+\sqrt{w^2-4}\,\right]
        \coth\left(\mu\,\mathrm{Arcosh}\frac{w}{2}\right)} 
      & : w>2
      % \end{cases}
    \end{array}
  \right.
\end{equation}
inferred from trigonometric/hyperbolic representations of Chebyshev
polynomials.  For large $\mu$ the functions $r_\mu(w)$ are much smoother than
the polynomials $S_\mu(w)$. Standard methods with initial values from the
analytic solution for $\mu\to\infty$ derived in Sec.~\ref{sec:second} below
work quite well.

%%%%%%%%%%%%%%%%%%%%%%%%%%%%%%%%%%%%%%%%%%%%%%%
\subsection{First-order transition}\label{sec:first}
%%%%%%%%%%%%%%%%%%%%%%%%%%%%%%%%%%%%%%%%%%%%%%%
In the limit $\tau\to0$ at $\mu<\infty$, the solution (\ref{eq:36}) acquires a
linear a cusp as shown in Fig.~\ref{fig:crossover}(a):
\begin{equation}\label{eq:20} 
  w=  
  \left\{ \begin{array}{ll}
      % \begin{cases}
      t_0 -1& : t\leq t_0 \\ t-1 & : t\geq t_0    
      % \end{cases}
  \end{array}    \right.
  \qquad (\tau=0),
\end{equation}
as the growth parameter $t$ increases across the transition value,
\begin{equation}\label{eq:21} 
t_0\doteq 1+2\cos\!\left(\frac{\pi}{\mu+1}\right)\qquad (\tau=0).
\end{equation}
It describes a discontinuous phase transition between a pure coil at $t<t_0$
and a pure helix at $t>t_0$ in a sense that requires some explanations
(Sec.~\ref{sec:ord-dis}). Discontinuities are manifest in the order parameter
and the entropy.  The latter is associated with a latent heat.

%%%%%%%%%%%%%%%%%%%%%%%%%%%%%%%%%%%%%%%%%%%%%%%
\subsection{Second-order transition}\label{sec:second}
%%%%%%%%%%%%%%%%%%%%%%%%%%%%%%%%%%%%%%%%%%%%%%%
In the limit $\mu\to\infty$ at $\tau>0$ the solution (\ref{eq:36}) acquires a
quadratic cusp at
\begin{equation}\label{eq:25} 
  t_\mathrm{c}\doteq \frac{3}{1+\tau}\qquad (\mu=\infty).
\end{equation}
Performing that limit in (\ref{eq:54}) yields
\begin{equation}\label{eq:55} 
r_\infty(w)=\frac{1}{2}\left[w+\sqrt{w^2-4}\,\right],\quad w\geq2.
\end{equation}
The resulting analytic solution then reads
\begin{equation}\label{eq:17}
w=\left\{ \begin{array}{ll} 2 & : 0\leq t\leq t_\mathrm{c} \\ 
t-1+{\displaystyle \frac{t\tau}{\lambda}} & : t>t_\mathrm{c}
\end{array} \right. \qquad (\mu=\infty),
\end{equation}
\begin{equation}\label{eq:51} 
\lambda\doteq \frac{1}{2}\left[t-1+\sqrt{(t+1)(t-3)+4t\tau}\,\right],
\end{equation}
and is graphically represented in Fig.~\ref{fig:crossover}(b).
The singularity at $t_\mathrm{c}^+$, 
\begin{equation}\label{eq:52} 
w=2+%\theta(t-t_\mathrm{c})
\left[\frac{t_0(t-t_\mathrm{c})}{t_\mathrm{c}(t_0-t_\mathrm{c})}\right]^2
+\mathcal{O}\big((t-t_\mathrm{c})^3\big)
\end{equation}
with $t_0=3$ for $\mu=\infty$ represents a continuous transition between a
coil phase $(t<t_\mathrm{c})$ and a helix phase $(t>t_\mathrm{c})$ with
helical ordering subject to thermal fluctuations.  Expression (\ref{eq:52}) is
to be interpreted as an asymptotic expansion with coefficients that diverge as
$\tau\to0$.  In that limit the cusp turns linear as in (\ref{eq:20}).

%%%%%%%%%%%%%%%%%%%%%%%%%%%%%%%%%%%%%%%%%%%%%%%
%
\section{Order and disorder}\label{sec:ord-dis}
%
%%%%%%%%%%%%%%%%%%%%%%%%%%%%%%%%%%%%%%%%%%%%%%%
Helix means order and coil means disorder, clearly.  However, both attributes
can be looked at from different angles and a more comprehensive picture
emerges.  In the following we investigate several thermodynamic quantities,
derived from the free energy (\ref{eq:11}) as functions of the experimentally
controllable growth parameter $t$ at fixed values of the other two parameters
$\tau$ and $\mu$.

Each quantity will illuminate the competition between order and disorder from
a somewhat different vantage point.  All are functions of $w_1(t,\tau)$, which
depends on the solution (\ref{eq:36}) of (\ref{eq:9}) via (\ref{eq:10}).  The
analytic expression in the limit $\tau\to0$ for $\mu<\infty$ as inferred from
(\ref{eq:20}) reads
\begin{equation}\label{eq:22} 
w_1=\left\{\begin{array}{ll}
{\displaystyle \frac{t}{t_0-t}} & : t< t_0 \\ \infty & : t\geq t_0
\end{array} \right.\qquad (\tau=0),
\end{equation}
and the analytic result in the limit $\mu\to\infty$ at $\tau>0$ as inferred
from (\ref{eq:17}) becomes
  \begin{eqnarray} \label{eq:39}
    w_{1}=
    \left\{\begin{array}{ll}
%    \begin{cases} 
      {\displaystyle \frac{t}{t_0-t}} 
      & : t < t_\mathrm{c} \\  \rule[-2mm]{0mm}{8mm}
     {\displaystyle \frac{\lambda}{\tau}} 
     & : t \geq t_\mathrm{c} \\
%    \end{cases} 
   \end{array}\right.
   \qquad (\mu=\infty).
  \end{eqnarray}

%%%%%%%%%%%%%%%%%%%%%%%%%%%%%%%%%%%%%%%%%%%%%%%
\subsection{Helicity and entropy}\label{sec:heli-ent}
\begin{figure}[b]
  \begin{center}
\includegraphics[width=47mm]{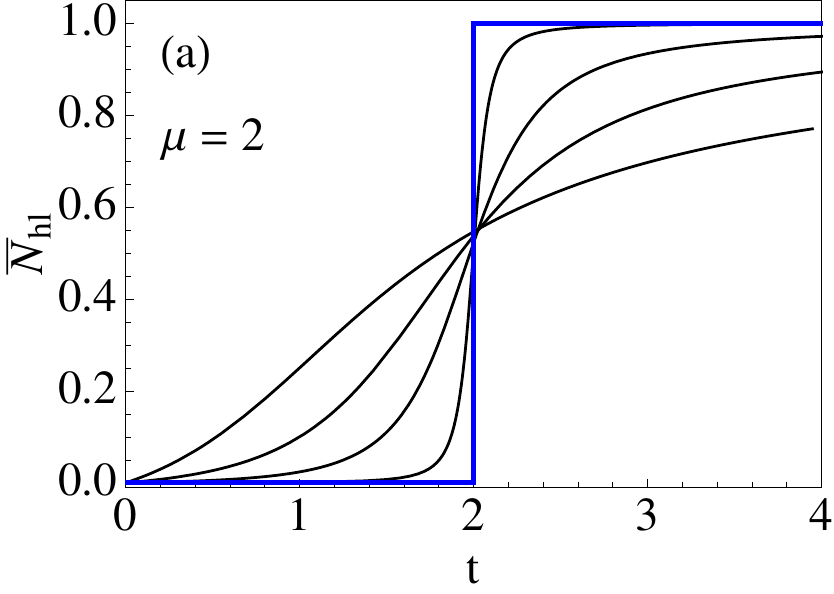}\hspace*{3mm}\includegraphics[width=47mm]{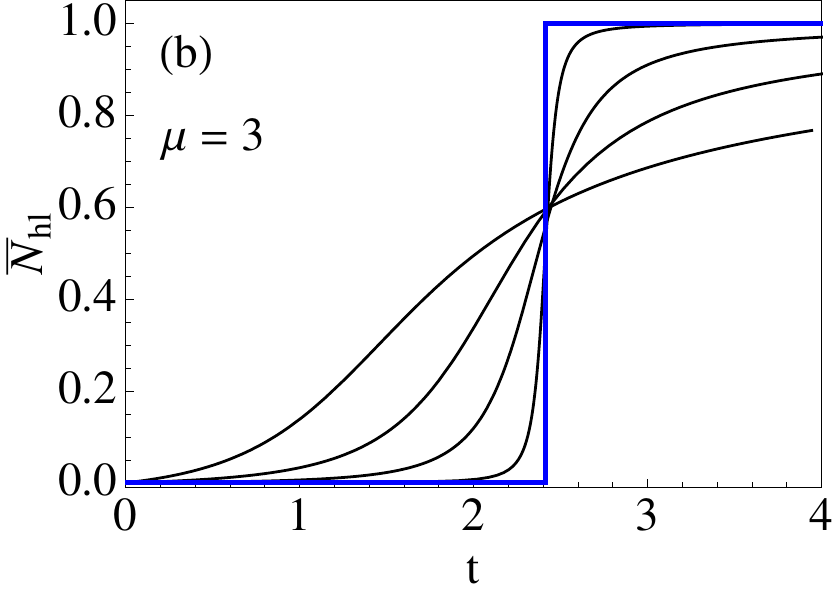}
\end{center}
\caption{(Color online) Helicity $\bar{N}_\mathrm{hl}$ versus growth parameter
  $t$ at values $\tau=1, 0.25, 0.05, 0.0025$ (thin lines) and $\tau=0$ (thick
  line) of the nucleation parameter for (a) $\mu=2$ and (b) $\mu=3$.}
  \label{fig:helicity1}
\end{figure}

\begin{figure}[b]
  \begin{center}
\includegraphics[width=47mm]{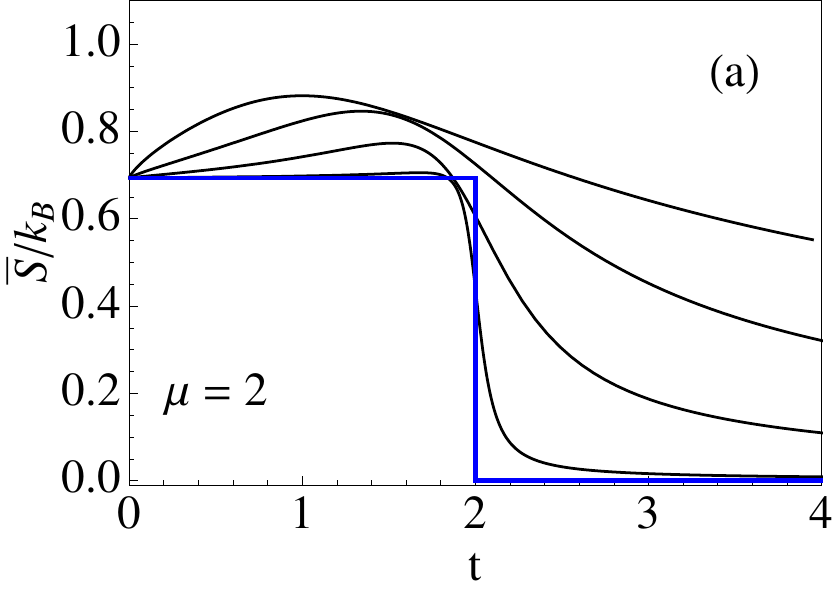}\hspace*{3mm}\includegraphics[width=47mm]{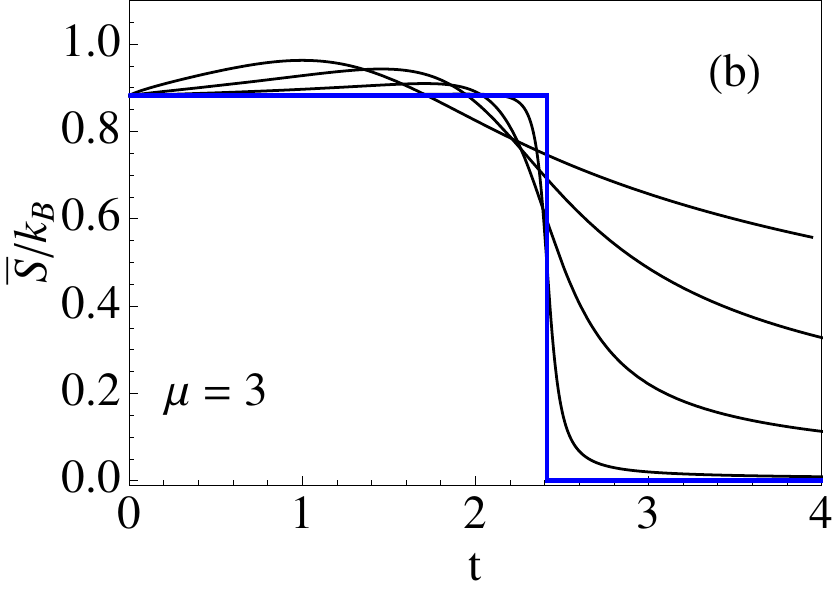}
\end{center}
\caption{(Color online) Entropy $\bar{S}/k_\mathrm{B}$ versus $t$ at $\tau=1,
  0.25, 0.05, 0.0025$ (thin lines) and $\tau=0$ (thick line) for (a) $\mu=2$
  and (b) $\mu=3$.}
  \label{fig:entropy1}
\end{figure}

%%%%%%%%%%%%%%%%%%%%%%%%%%%%%%%%%%%%%%%%%%%%%%%
We begin by considering the two thermodynamic functions that represent order
and disorder most directly: helicity (\ref{eq:14}),
\begin{equation}\label{eq:26}
\bar{N}_\mathrm{hl}=1-
\frac{t}{w_{1}(1+w_{1})}\frac{\partial w_{1}}{\partial t},
\end{equation}
and entropy (\ref{eq:12}),
\begin{eqnarray}\label{eq:27} 
  \frac{\bar{S}}{k_\mathrm{B}} &=
  \ln\Big(1+w_{1}^{-1}\Big) % \nonumber \\ &\hspace{-0mm}
  +\frac{1}{w_{1}(1+w_{1})}
  \left[t\ln t\frac{\partial w_{1}}{\partial t}+
    \tau\ln\tau\frac{\partial w_{1}}{\partial \tau}\right].
\end{eqnarray}
In Figs.~\ref{fig:helicity1} and \ref{fig:entropy1} we show the dependence of
helicity and entropy on the growth parameter $t$ at fixed nucleation parameter $\tau$ (five curves)
and range parameter $\mu$ (two panels).

%%%%%%%%%%%%%%%%%%%%%%%%%%%%%%%%%%%%%%%%%%%%%%%
%%%%%%%%%%%%%%%%%%%%%%%%%%%%%%%%%%%%%%%%%%%%%%%

%%%%%%%%%%%%%%%%%%%%%%%%%%%%%%%%%%%%%%%%%%%%%%%
%%%%%%%%%%%%%%%%%%%%%%%%%%%%%%%%%%%%%%%%%%%%%%%

At finite cooperativity $(\tau>0)$ the helicity crosses over from a low to a
high value near $t_0$.  The rise in helicity becomes sharper with increasing
cooperativity and turns into a step discontinuity in the limit $\tau\to0$.
Analytically, expression (\ref{eq:26}) with (\ref{eq:22}) substituted yields
\begin{equation}\label{eq:30} 
\bar{N}_\mathrm{hl}=\theta\!\left(t-t_0\right)\qquad (\tau=0).
\end{equation}

While order as reflected in the helicity increases monotonically with $t$, the
disorder as reflected in the entropy is not monotonically decreasing.  It
shows a shallow maximum at $t\simeq1$ separate from the shoulder at $t\simeq
t_0$.  The reason for this difference is that there is only one source of
order -- helical links -- but two sources of disorder: disorder inside coil
segments and disorder in the sequence of helical/coil segments of diverse
lengths.  It is the first source of disorder that produces the shoulder and
the second source that produces the shallow maximum.
The Zimm-Bragg case $\mu=1$ is pathological in this respect.
It produces coil segments without internal entropy as noted and commented on 
before \cite{NRV07}.

As $\tau\to0$, the segments grow larger in size and become fewer in numbers
(see Sec.~\ref{sec:seg-coi-hel} below). This reduces disorder of the second
kind.  At infinite cooperativity the entropy turns into a step discontinuity
of height and location.  Expression (\ref{eq:27}) with (\ref{eq:22})
substituted becomes
\begin{equation}\label{eq:31} 
\frac{\bar{S}}{k_B}=\theta\!\left(t_0-t\right)\ln t_0\qquad (\tau=0).
\end{equation}
This discontinuity signals the presence of a latent heat (see
Sec.~\ref{sec:heat-late} below).

\begin{figure}[b]
  \begin{center}
\includegraphics[width=47mm]{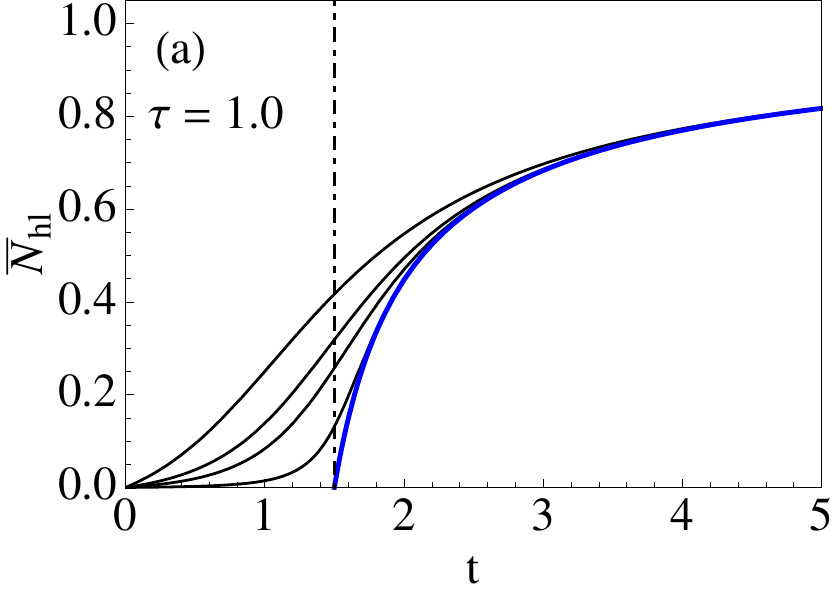}\hspace*{3mm}\includegraphics[width=47mm]{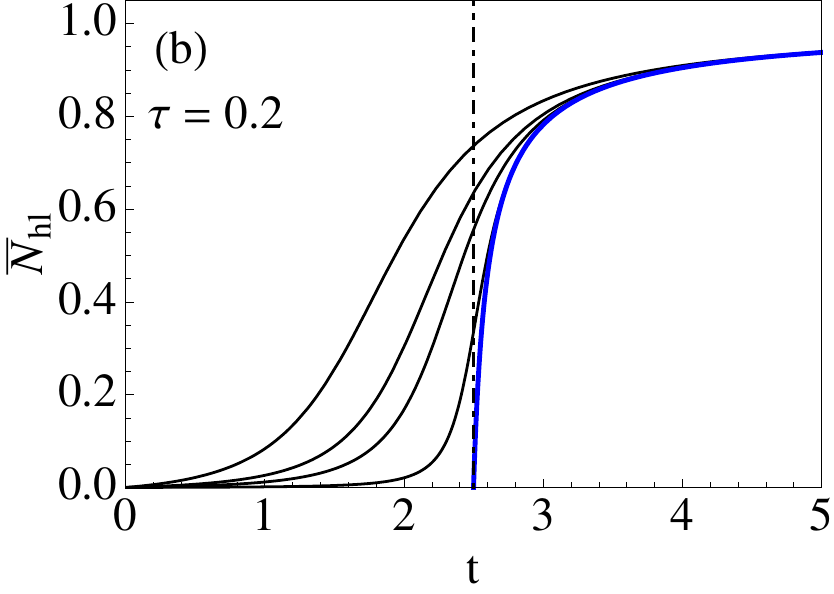}
\end{center}
\caption{(Color online) Helicity $\bar{N}_\mathrm{hl}$ versus growth parameter
  $t$ at cooperativity (a) $\tau=1.0$ and (b) $\tau=0.2$ for $\mu=2, 3, 4, 9$
  (thin curves from top down) and $\mu=\infty$ (thick curve).  The dot-dashed
  lines marks $t_\mathrm{c}$.}
  \label{fig:helicity2}
\end{figure}

\begin{figure}[b]
  \begin{center}
\includegraphics[width=47mm]{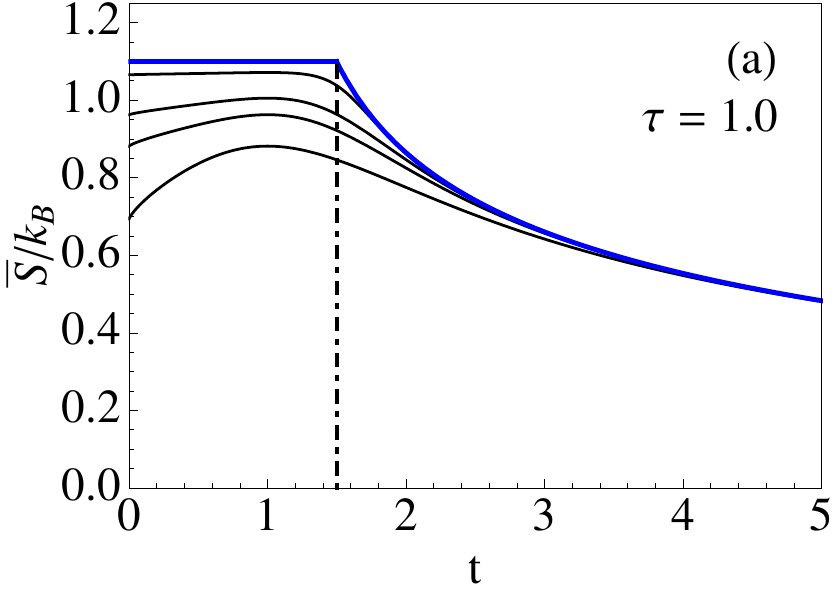}\hspace*{3mm}\includegraphics[width=47mm]{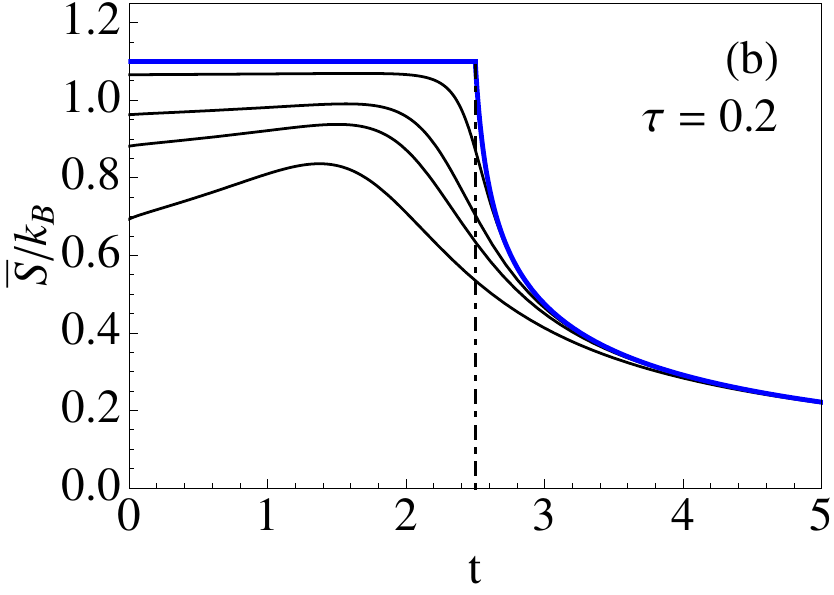}
\end{center}
\caption{(Color online) Entropy $\bar{S}/k_\mathrm{B}$ versus $t$ at (a)
  $\tau=1.0$ and (b) $\tau=0.2$ for $\mu=2, 3, 4, 9$ (thin curves from bottom
  up) and $\mu=\infty$ (thick curve).  The dot-dashed line marks
  $t_\mathrm{c}$.}
  \label{fig:entropy2}
\end{figure}

Next we investigate the same measures of order and disorder as functions of
$t$ at fixed $\tau=1.0$ (low cooperativity) or $\tau=0.2$ (high cooperativity)
for increasing numbers $\mu$ of coil states per residue including the limit
$\mu\to\infty$.  Our results are shown in Figs.~\ref{fig:helicity2} and
\ref{fig:entropy2}.  The crossover behavior for small $\mu$ turns into a
continuous order-disorder transition as $\mu\to\infty$.

%%%%%%%%%%%%%%%%%%%%%%%%%%%%%%%%%%%%%%%%%%%%%%%
%%%%%%%%%%%%%%%%%%%%%%%%%%%%%%%%%%%%%%%%%%%%%%%

%%%%%%%%%%%%%%%%%%%%%%%%%%%%%%%%%%%%%%%%%%%%%%%
%%%%%%%%%%%%%%%%%%%%%%%%%%%%%%%%%%%%%%%%%%%%%%%

With $\mu$ increasing, the internal source of disorder in coil segments gains
dominance over the entropy of mixing between coil and helix segments.  The
shoulder in $\bar{S}/k_\mathrm{B}$ becomes flatter, higher, and sharper.
In the limit $\mu\to\infty$ at $t<t_\mathrm{c}$, the helicity approaches zero
identically and the entropy approaches the value $\bar{S}/k_\mathrm{B}=\ln 3$,
independent of $\tau$.  Disorder defeats order hands down.

The helix phase at $t>t_\mathrm{c}$, by contrast, remains a battleground
between ordering and disordering tendencies.  Both the helicity and the
entropy expressions,
\begin{equation}\label{eq:2}
  \bar{N}_\mathrm{hl}=1-\frac{t\tau}{\lambda(2\lambda+1-t)},
\end{equation}
\begin{eqnarray}\label{eq:56} 
  \frac{\bar{S}}{k_\mathrm{B}}=\ln\left(1+\frac{\tau}{\lambda}\right) &+
  \frac{t\tau\ln t}{\lambda(2\lambda+1-t)} % \nonumber \\ &
  +\frac{\tau\ln\tau}{\lambda+\tau}
  \left[\frac{t\tau}{\lambda(2\lambda+1-t)}-1\right],
\end{eqnarray}
have linear cusps at $t_\mathrm{c}$ with slopes that diverge in the limit
$\tau\to0$. The leading critical singularities at $t_\mathrm{c}^+$ are
\begin{equation}\label{eq:57} 
  \bar{N}_\mathrm{hl}
  =\frac{2t_0}{t_\mathrm{c}}\frac{t-t_\mathrm{c}}{(t_0-t_\mathrm{c})^2}
  +\mathcal{O}\big((t-t_\mathrm{c})^2\big),
\end{equation}
\begin{eqnarray}\label{eq:58} 
  \frac{\bar{S}}{k_\mathrm{B}}
  &=\ln t_0-2\frac{t-t_\mathrm{c}}{t_0-t_\mathrm{c}}
  \left[\frac{\ln t_\mathrm{c}}{t_0-t_\mathrm{c}} +
    \frac{\ln(t_0-t_\mathrm{c})}{t_\mathrm{c}}\right] 
  % \\ \nonumber &\hspace{50mm}
  +\mathcal{O}\big((t-t_\mathrm{c})^2\big).
\end{eqnarray}
With the growth parameter increasing from $t_\mathrm{c}$ the helicity steeply
rises from zero and gradually bends over toward its saturation value whereas
the entropy steeply descends from a high value and gradually approaches zero.
The plots suggest that cooperativity impedes the onset of ordering, yet
assists the quick rise of ordering once it has set in.

%%%%%%%%%%%%%%%%%%%%%%%%%%%%%%%%%%%%%%%%%%%%%%%
\subsection{Segments of coil and helix}\label{sec:seg-coi-hel}
%%%%%%%%%%%%%%%%%%%%%%%%%%%%%%%%%%%%%%%%%%%%%%%
Further insight into how helical ordering grows during the crossover or near
the transition point between conformations can be gained from the two
quantities (\ref{eq:15}) and (\ref{eq:16}), representing, respectively, the
density and average length of segments in one or the other conformation.
Coil segments alternate with helix segments.  Hence they come in equal
numbers.  However, their average lengths vary independently with $t$.
Parametric representations can be constructed as before. We use (\ref{eq:26})
and
\begin{equation}\label{eq:32} 
  \bar{N}_\mathrm{seg}
  =-\frac{\tau}{w_{1}(1+w_{1})}\frac{\partial w_{1}}{\partial \tau}.
\end{equation}

\begin{figure}[b]
  \begin{center}
\includegraphics[width=47mm]{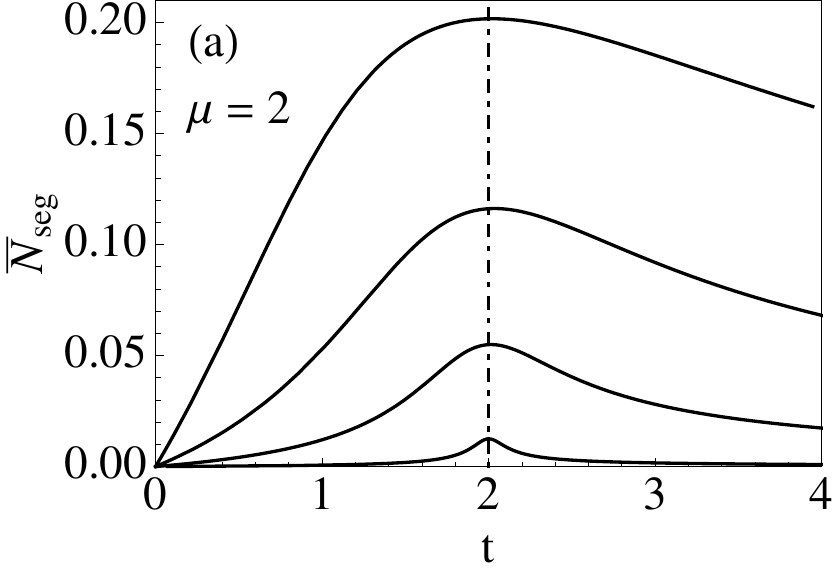}\hspace*{3mm}\includegraphics[width=47mm]{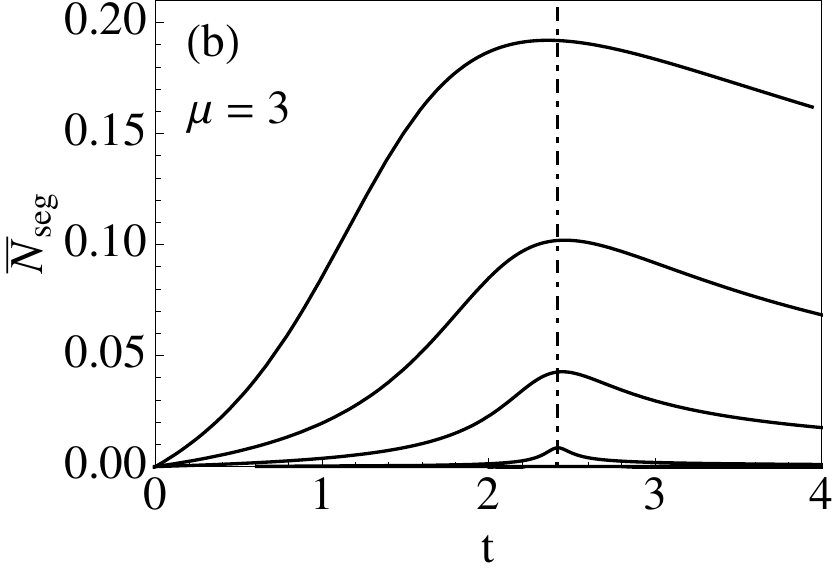}
\end{center}
\caption{Density of segments, $\bar{N}_\mathrm{seg}$, versus $t$ at $\tau=1$,
  $0.25$, $0.05$, $0.0025$ (from top down) for (a) $\mu=2$ and (b) $\mu=3$.
  The dot-dashed line marks $t_0$.}
  \label{fig:nseg-fot}
\end{figure}

\begin{figure}[t]
  \begin{center}
\includegraphics[width=47mm]{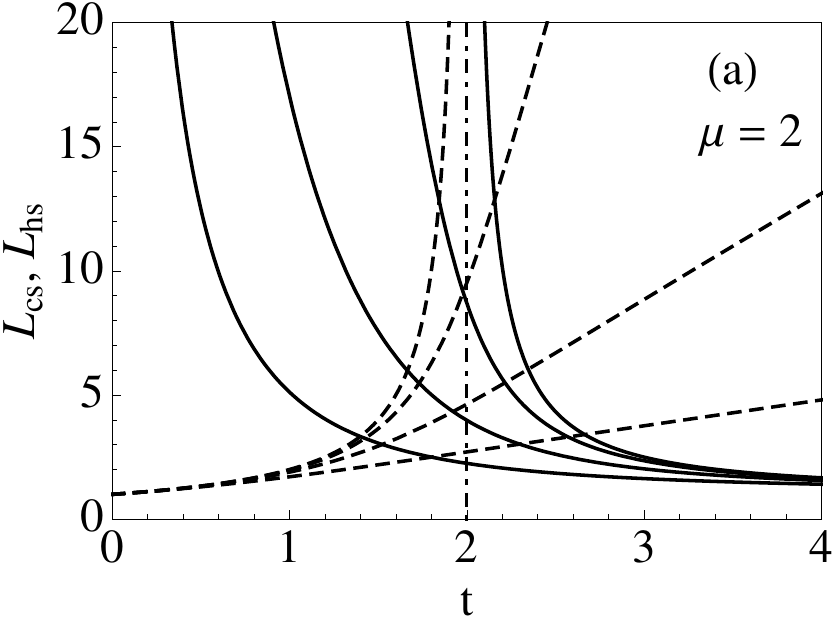}\hspace*{3mm}\includegraphics[width=47mm]{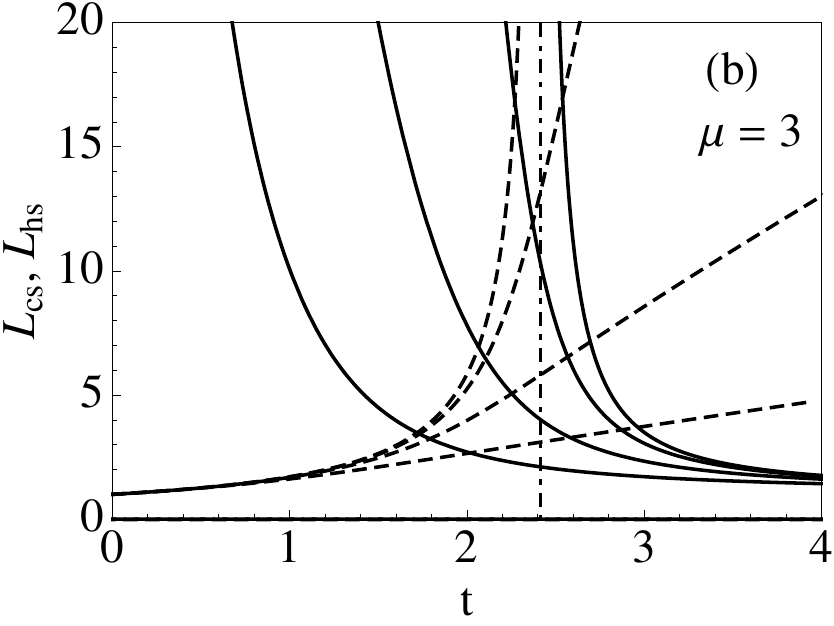}
\end{center}
\caption{Average length of coil segments, $L_\mathrm{cs}$ (solid lines), and
  helix segments, $L_\mathrm{hs}$ (dashed lines), versus $t$ at $\tau=1$,
  $0.25$, $0.05, 0$ (from bottom up) for (a) $\mu=2$ and (b) $\mu=3$. The
  dot-dashed line marks $t_0$.}
  \label{fig:lhs-lcs-fot}
\end{figure}

We first examine the $t$-dependence of $\bar{N}_\mathrm{seg}$,
$L_\mathrm{cs}$, and $L_\mathrm{hs}$ near the first-order transition that
takes place at $t_0$ in the limit $\tau\to0$.  Our results for $\mu=2,3$ are
shown in Figs.~\ref{fig:nseg-fot} and \ref{fig:lhs-lcs-fot}.
%%%%%%%%%%%%%%%%%%%%%%%%%%%%%%%%%%%%%%%%%%%%%%%
%%%%%%%%%%%%%%%%%%%%%%%%%%%%%%%%%%%%%%%%%%%%%%%
We observe that the density of segments is near zero at small $t$.  Here the
system is strongly coil-like.  The segments grow in numbers with $t$
increasing.  They become most numerous at $t_0$, where ordering and
disordering tendencies compete evenly.  The density of segments becomes
smaller again as $t$ further increases into the stability regime of the helix
conformation.  The maximum of $\bar{N}_\mathrm{seg}$ at $t_0$ strongly depends
on $\tau$.  In the limit $\tau\to0$ we have $\bar{N}_\mathrm{seg}\equiv0$,
which means that, in a macroscopic system, the number of segments grows more
slowly (if at all) than the number of residues.

%%%%%%%%%%%%%%%%%%%%%%%%%%%%%%%%%%%%%%%%%%%%%%%
%%%%%%%%%%%%%%%%%%%%%%%%%%%%%%%%%%%%%%%%%%%%%%%

Unsurprisingly, $L_\mathrm{cs}$ decreases and $L_\mathrm{hs}$ increases
monotonically with $t$.  As expected, both variations are enhanced by
cooperativity.  Most interesting is the limit $\tau\to0$.  The exact
expressions for the average lengths of helix segments at $t<t_0$ and coil
segments at $t>0$ read
\begin{eqnarray}\label{eq:33} 
  L_\mathrm{hs} &= \frac{t_0}{t_0-t},
  \\ \label{eq:34} 
  L_\mathrm{cs} &= \frac{t}{(1+t)(3-t)}
  \Bigg[\frac{2(\mu+1)}{r_\mu(t-1)} +2\mu
  r_\mu(t-1)-(2\mu+1)(t-1)\Bigg],
\end{eqnarray}
respectively, with $r_\mu(w)$ from (\ref{eq:54}).  Both expressions diverge
$\propto|t-t_0|^{-1}$ as $t$ approaches $t_0$ from opposite sides and then
stay infinite.

These results tell us that the macrostate with zero helicity and saturated
entropy at $t<t_0$ still contains helix segments albeit only in numbers that
do not add up to a nonzero density but still produce a well-defined average
size.  They coexist with an equal number of coil segments of macroscopic
lengths.
Conversely, the macrostate of zero entropy (per residue) and saturated
helicity at $t>t_0$ is not a single helical domain.  Here helix segments of
short average length in numbers that amount to zero density coexist with an
equal number of helix segments of macroscopic lengths.

\begin{figure}[htb]
  \begin{center}
\includegraphics[width=47mm]{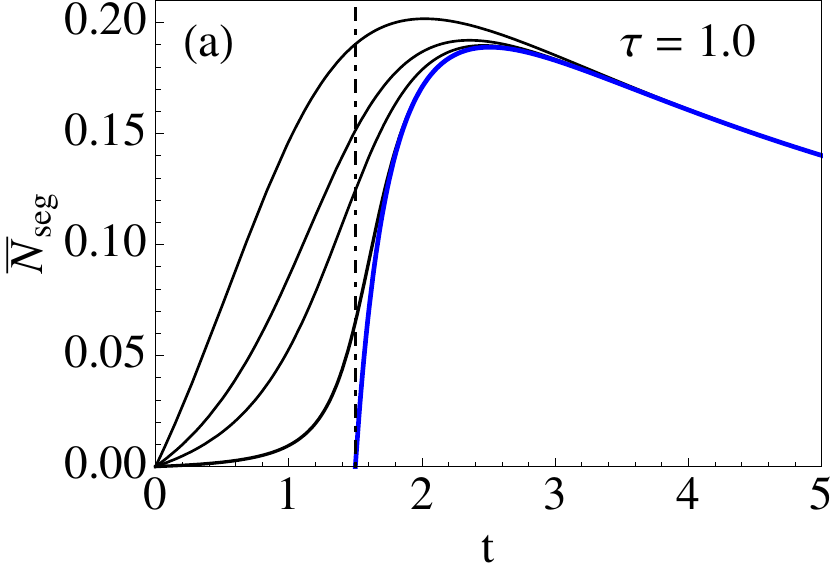}\hspace*{3mm}\includegraphics[width=47mm]{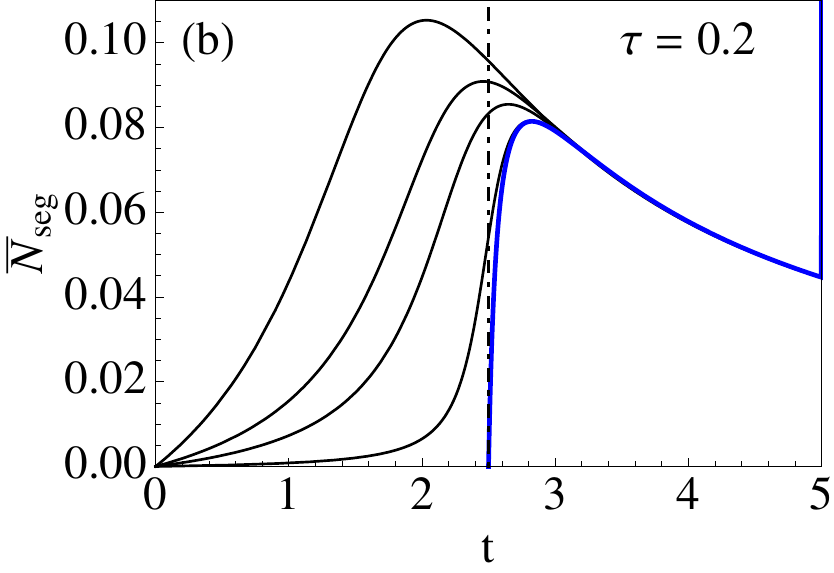}
\end{center}
\caption{(Color online) Density of segments, $\bar{N}_\mathrm{seg}$, versus
  $t$ at (a) $\tau=1.0$ and (b) $\tau=0.2$ for $\mu=2, 3, 4, 9$ (thin curves
  lines left to right) and $\mu=\infty$ (thick curve). The dot-dashed line
  marks $t_\mathrm{c}$.}
  \label{fig:nseg-nu}
\end{figure}

\begin{figure}[htb]
  \begin{center}
\includegraphics[width=47mm]{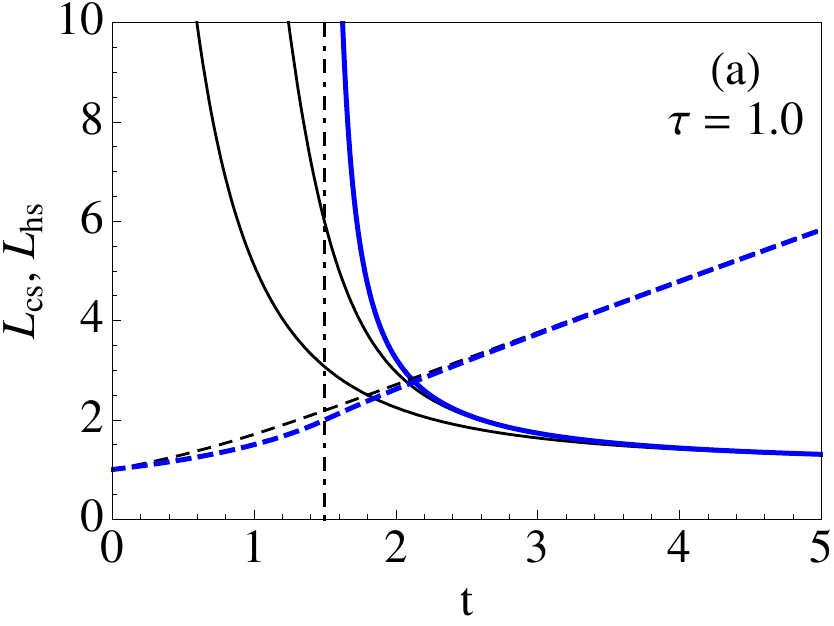}\hspace*{3mm}\includegraphics[width=47mm]{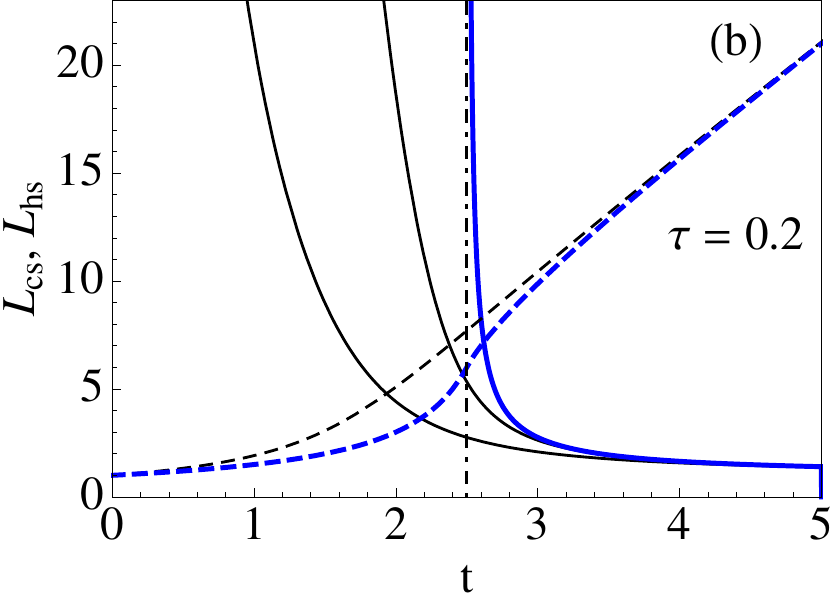}
\end{center}
\caption{(Color online) Average length of coil segments, $L_\mathrm{cs}$
  (solid lines), and helix segments, $L_\mathrm{hs}$ (dashed lines), for
  $\mu=2$ (thin lines) and $\mu=\infty$ (thick lines) versus $t$ at (a)
  $\tau=1.0$ and (b) $\tau=0.2$. The dot-dashed line marks $t_\mathrm{c}$.}
  \label{fig:lhs-lcs-nu}
\end{figure}

A different picture emerges near the second-order transition at
$t=t_\mathrm{c}$ in the limit $\mu\to\infty$.  Results for the density of
segments at high and low cooperativity are shown in Fig.~\ref{fig:nseg-nu} and
results for their average lengths in Fig.~\ref{fig:lhs-lcs-nu}.  This includes
numerical results for $\mu<\infty$ and analytical results for $\mu=\infty$.
%%%%%%%%%%%%%%%%%%%%%%%%%%%%%%%%%%%%%%%%%%%%%%%
%%%%%%%%%%%%%%%%%%%%%%%%%%%%%%%%%%%%%%%%%%%%%%%
%%%%%%%%%%%%%%%%%%%%%%%%%%%%%%%%%%%%%%%%%%%%%%%
%%%%%%%%%%%%%%%%%%%%%%%%%%%%%%%%%%%%%%%%%%%%%%%
The density of segments vanishes identically in the coil phase
$(t<t_\mathrm{c})$ and then rises in a linear cusp to a smooth maximum in the
helix phase $(t>t_\mathrm{c})$ :
\begin{eqnarray}\label{eq:57} 
  \bar{N}_\mathrm{seg} 
  =\frac{\tau}{\lambda+\tau}\left[1-\frac{t\tau}{\lambda(2\lambda+1-t)}\right]
  =\frac{2t_0}{t_\mathrm{c}}\frac{t-t_\mathrm{c}}{t_0-t_\mathrm{c}}
  +\mathcal{O}\big((t-t_\mathrm{c})^2\big).
\end{eqnarray}
The average length of coil segments in the helix phase,
\begin{eqnarray}\label{eq:23}
  \hspace*{-5mm}L_\mathrm{cs} 
  =\frac{t(\lambda+\tau)}{\lambda(2\lambda+1-t)-t\tau} 
  =\frac{t_\mathrm{c}}{2}\frac{t_0-t_\mathrm{c}}{t-t_\mathrm{c}}
  +\frac{2t_\mathrm{c}^2-9t_\mathrm{c}+15}{4(t_0-t_\mathrm{c})} 
  +\mathcal{O}(t-t_\mathrm{c}),
\end{eqnarray}
diverges at $t_\mathrm{c}^+$ and remains infinite in the coil phase.  The
average length of helix segments, by contrast, remains finite in both phases,
\begin{equation}\label{eq:59} 
  L_\mathrm{hs}=\left\{ \begin{array}{ll} 
      {\displaystyle \frac{t_0}{t_0-t}} 
      & : t\leq t_\mathrm{c}, \\ \rule[-2mm]{0mm}{8mm}
      {\displaystyle 1+\frac{\lambda}{\tau}} 
      & : t\geq t_\mathrm{c}, 
    \end{array} \right.
\end{equation}
where again, $t_0=3$ for $\mu=\infty$ in (\ref{eq:57})-(\ref{eq:59}).  The
graph of $L_\mathrm{hs}$ is continuous and smooth at $t_\mathrm{c}$.  The
singularity is of higher order.  Only in the helix phase does the shape of the
curve depend on $\tau$.

The most striking feature in the data shown concerns the helix segments.
Unlike in the case of the first-order transition, the ordered phase near
$t_\mathrm{c}$ supports a significant density of coil and helix segments of
comparable finite size.  The average length of helix segments depends only
weakly on $\mu$ and only moderately on $\tau$, in strong contrast to the
average length of coil segments, which exhibits strong dependences on both
parameters.

%%%%%%%%%%%%%%%%%%%%%%%%%%%%%%%%%%%%%%%%%%%%%%%
\subsection{Heat capacity and latent heat}\label{sec:heat-late}
\begin{figure}[htb]
  \begin{center}
\includegraphics[width=47mm]{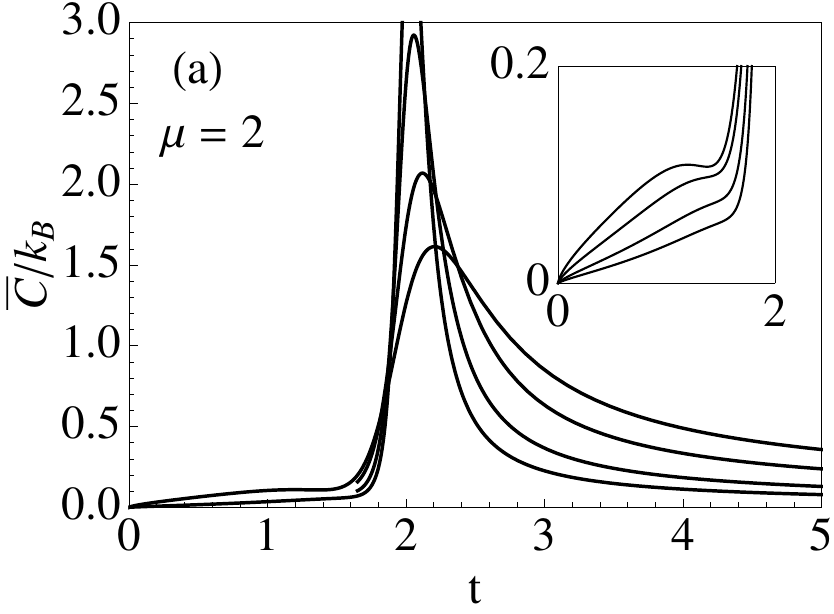}\hspace*{3mm}\includegraphics[width=47mm]{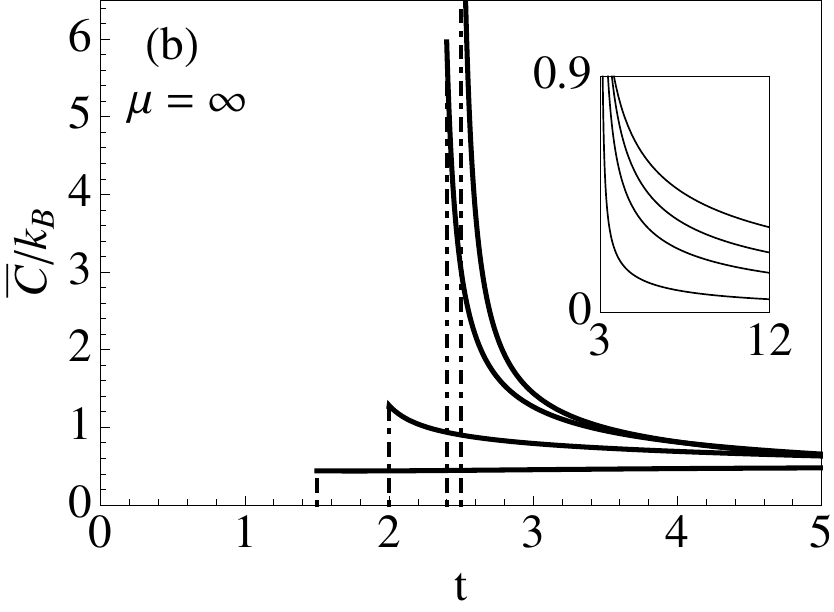}
\end{center}
\caption{Heat capacity $\bar{C}/k_\mathrm{B}$ versus $t$ for (a) $\mu=2$ at
  $\tau=0.05$, $0.025$, $0.01$, $0.005$ (from top down on the right in the
  main plot and from bottom up in the inset) and (b) $\mu=\infty$ at
  $\tau=1.0$, $0.5$, $0.25$, $0.2$ (from bottom up in main plot) and
  $\tau=0.2$, $0.1$, $0.05$, $0.01$ (from top down in the inset). The
  dot-dashed lines mark $t_c$ for given $\tau$.}
  \label{fig:he-ca1}
\end{figure}

%%%%%%%%%%%%%%%%%%%%%%%%%%%%%%%%%%%%%%%%%%%%%%%
The heat capacity, $\bar{C}\doteq T(\partial\bar{S}/\partial
T)_{\epsilon_\mathrm{n},\epsilon_\mathrm{g}}$, illuminates the competition
between order and disorder from yet a different angle.  From (\ref{eq:27}) we
derive
\begin{eqnarray}\label{eq:35} \fl
  \frac{\bar{C}}{k_B} 
  = \frac{2w_1+1}{w_1^2(1+w_1)^2}\left[ t\ln t\frac{\partial w_1}{\partial t} 
    +\tau\ln\tau\frac{\partial w_1}{\partial\tau}\right]^2 
  % \nonumber \\ &\hspace{-10mm} 
  -\frac{1}{w_1(1+w_1)}
  \left[t(\ln t)^2\frac{\partial w_1}{\partial t}\right.
  \nonumber \\ \fl % &\hspace{15mm} 
  \hspace{10mm} 
  \left. +\tau(\ln\tau)^2\frac{\partial w_1}{\partial\tau}
  +(t\ln t)^2\frac{\partial^2 w_1}{\partial t^2} 
  +(\tau\ln\tau)^2\frac{\partial^2 w_1}{\partial\tau^2} 
  % \nonumber \\ &\hspace{15mm}\left.
    +2(t\ln t)(\tau\ln\tau)
    \frac{\partial^2w_1}{\partial t\partial\tau}\right].
\end{eqnarray}
Figure~\ref{fig:he-ca1} shows the dependence of the heat capacity on the
growth parameter for the case $\mu=2$ at moderate to high cooperativity and
for the case $\mu=\infty$ over a wider range of cooperativity.

%%%%%%%%%%%%%%%%%%%%%%%%%%%%%%%%%%%%%%%%%%%%%%%
%%%%%%%%%%%%%%%%%%%%%%%%%%%%%%%%%%%%%%%%%%%%%%%

At $t<t_0=2$ in panel (a) we observe a weak signal that is associated with the
entropy caused by alternating coil and helix segments as discussed previously.
This contribution fades away at high cooperativity (best seen in the inset) as
the density of segments diminishes.
The peak at $t\gtrsim t_0$, on the other hand, is associated with the entropy
inside coil segments.  With increasing cooperativity, this contribution grows
in a more and more narrow range at $t_0$.
Similar structures have been obtained in recent Monte Carlos simulations, albeit upon variation of temperature and in a somewhat different scenario \cite{CZD04}.

In the limit $\tau\to0$ for $\mu<\infty$, where the coil-helix crossover
sharpens into a first-order transition at $t_0$, the heat capacity approaches
zero everywhere except at the transition point, where it diverges and
produces, via (\ref{eq:31}), a latent heat of magnitude $\epsilon_\mathrm{g}$.
Conversely, in the limit $\mu\to\infty$ at $\tau>0$, where the coil-helix
crossover turns into a second-order transition at $t_\mathrm{c}$, the heat
capacity approaches zero in the coil conformation and remains nonzero in the
helix conformation as shown in panel (b).  When the transition changes from
second to first order when $\tau\to0$ for $\mu=\infty$, implying
$t_\mathrm{c}\to t_0=3$, the heat capacity throughout the helix conformation
approaches zero as illustrated in the inset.

%%%%%%%%%%%%%%%%%%%%%%%%%%%%%%%%%%%%%%%%%%%%%%%
%
\section{Conclusion and outlook}\label{sec:con-out}
%
%%%%%%%%%%%%%%%%%%%%%%%%%%%%%%%%%%%%%%%%%%%%%%%
We have launched this project mainly for the purpose of interpreting (ongoing
and projected) experiments on pHLIP.  In this first of three stages of
analysis we have constructed a microscopic model for the pH-driven coil-helix
conformational change of a long polypeptide adsorbed to a water-lipid
interface.
We have employed a methodology that facilitates the exact statistical
mechanical analysis of our model.  The three model parameters $t, \tau, \mu$
have settings for which the conformation changes either in a crossover, a
first-order transition, or a second-order transition.

We have carried out the analysis to the extent needed for a discussion of the
sources and agents of order and disorder.  Our results include the
$t$-dependence of the helicity (order parameter), the average numbers and the
average lengths of helix and coil segments, the entropy, and the heat
capacity.  The behavior of these quantities near the continuous or the
discontinuous transition has been given special attention.
We have plotted all quantities versus $t$ at constant $\tau$ and $\mu$ for a
reason.  The experimentally relevant processes for which we use our model will
primarily involve variations of the growth parameter $t$.  These variations
are caused by changes in pH.  The targeted peptides, adsorbed to the
water-lipid interface, include side chains that are strongly hydrophobic
(e.g. Leu) and side chains that are negatively charged (e.g. Asp, Glu).

A drop in pH leads to the protonation of the negatively charged side chains
and, therefore, enhances the overall hydrophobicity.  The backbone of a coil
segment thus pushed past the lipid headgroups is now more likely to satisfy an
H-bond internally than externally.  The enthalpic cost for broken internal
H-bonds increases.  This cost is encoded in $t$.
Any increase in $t$ favors a growth of helix segments at the expense of coil
segments.  A rise in pH has the opposite effect.  The value of $t$ decreases.
Coil segments grow at the expense of helix segments.

The cooperativity parameter $\tau$, by contrast, is much less sensitive to a
change in pH.  In the nucleation process of coil segments from the helix
conformation, for example, the internal H-bonds are much more isolated from
environmental influences than are those at the border between coil segments
and helix segments.

At this point, our project has reached a fork, where natural continuations
point in two different directions and address the interests of somewhat
different audiences.  These continuations, already in the works, are outlined
as follows.

%%%%%%%%%%%%%%%%%%%%%%%%%%%%%%%%%%%%%%%%%%%%%%%
\subsection{Heterogeneous environment and short peptides}\label{sec:het-env}
%%%%%%%%%%%%%%%%%%%%%%%%%%%%%%%%%%%%%%%%%%%%%%%
In one continuation we begin by considering long polypeptides that are no
longer confined to a plane parallel to a flat water-lipid interface.  The
growth parameter $t$, which drives the conformational change, then becomes a
field $t(x)$ and acquires a profile that depends on the local medium.  Here
$x$ is a position coordinate in the direction perpendicular to the plane of
the membrane.  Such circumstances pose a serious challenge to any existing
model and its method of analysis.  However, the methodology used here is well
positioned in that respect.  It has already been proven (in different
applications \cite{janac2, mcut1, inharo}) to be adaptable to heterogeneous
environments.

The shape of the parameter field $t(x)$ will be determined by the availability
of polar molecules to satisfy external H-bonds along the backbone of the
peptide.  The dominant factor that shapes the field $t(x)$ will be the density
profile $\rho_\mathrm{w}(x)$ across the membrane, for which data from
experiments \cite{ALE+03} and simulations \cite{MBT08} are available.
Subdominant factors include electrostatic interactions and fluid-mechanical
properties of lipids.

From the analysis of our extended model emerge profiles for the densities of
free energy, enthalpy, entropy, and helicity of long polypeptides that
traverse the heterogeneous environment (ranging from polar to non-polar) along
some path that is subject to conformational constraints \cite{felop}.  These
profiles, in turn, will be interpreted as propensities for the statistical
mechanical behavior of short peptides in the same environment.

At this stage of the analysis, additional enthalpic and entropic effects
involving the side chains, the semi-fluid bilayer of lipid amphiphiles, and
the hydrogen-bonded network of $\mathrm{H_2O}$ molecules can be built into the
model.  The outcome are landscapes of free energy, enthalpy, entropy, and
helicity for short peptides of specific composition.
The free-energy landscapes in particular then set the stage for (i) a
theoretical study of the kinetics of trans-membrane insertion and exit of
pHLIP and other membrane peptides and (ii) a direct comparison with
experiments currently in progress that investigate the insertion/exit
processes of pHLIP via tryptophan fluorescence and the accompanying
conformational changes via circular dichroism spectroscopy.

This first continuation can also benefit from recent studies in the same area of research.
Not yest included in our modeling are effects related to torsion and tension, which are bound to be 
present in the heterogeneous membrane environment.
Experimental, computational, and analytic studies of force-extension and torque-twist 
characteristics and the associated steric constraints \cite{TP01, HSKN10, CL05, CL06} will be of 
great value for that purpose.
The kinetic modeling of pHLIP insertion while undergoing a conformational change will find 
valuable guidance from recent studies that have investigated the fluctuation properties of helical 
polymers in confined environments including narrow channels \cite{LBG04, YBG07, BYG10} 
and studies that have investigated the Brownian dynamics of polymers in the membrane 
environment \cite{RKSG11}.

%%%%%%%%%%%%%%%%%%%%%%%%%%%%%%%%%%%%%%%%%%%%%%%
\subsection{Extensions of analysis, model, and scope}\label{sec:c-s-m-e}
%%%%%%%%%%%%%%%%%%%%%%%%%%%%%%%%%%%%%%%%%%%%%%%
A second continuation focuses on the statistical mechanics of phase
transitions and critical singularities in the context of the microscopic model
presented in this work and extensions thereof.
It is well known that the presence of a phase transition at nonzero
temperature in a system that is, in some sense, one-dimensional requires
interactions of long-range to stabilize an ordered phase in the face of strong
thermal fluctuations.  In our model, which is truly microscopic and analyzed
exactly, this stabilizing agent comes in the form of quasiparticles that
extend over entire coil segments (hosts) or over parts thereof (hybrids).

In the context of the experiments that motivated this work we have examined
conformational changes driven by the control parameter $t$ at fixed $\tau,
\mu$ as reflected in just a few relevant quantities.  The phase transitions
that occur in the limits $\tau\to0$ (first-order) or $\mu\to\infty$ (second
order) produce different singularities in other quantities of no less interest
for the statistical mechanical analysis.
Such quantities of general interest include a mechanical response function, a
correlation length, and a correlation function.  The further analysis of
critical exponents and scaling laws is best presented in a more general
framework and along with model extensions that remain inside the reach of our
method of exact analysis.

In a final note, we should like to draw the reader's attention to a different
set of applications, for which our statistical mechanical model and its
extensions are likely to produce significant new insights. These applications
investigate the statistical mechanics and the dynamics of DNA melting (thermal
denaturation) \cite{PS66, Fish66, PB89, DPB93, TDP00, KMP00, KCG00, CH97,
COS02} or the loop formation in RNA \cite{EN11}.
Of particular interest is the loop exponent in the configurational entropy of loop formation 
\cite{Fish66, EN11}, which is frequently used as an adjustable parameter.
The further development of our project aims for the analytic calculation of loop exponents 
pertaining to realistic scenarios.
Discussions of and debates about crossovers, first-order transitions, and
second-order transitions are at the center of most of these studies.  The
transcription and adaptation of our methodology to this particular physics
context is already in progress.  The main challenge in the endeavor is the
extension of the self-avoiding random walk to three dimensions.

%%%%%%%%%%%%%%%%%%%%%%%%%%%%%%%%%%%%%%%%%%%%%%%
%
%\acknowledgments
\ack
%
%%%%%%%%%%%%%%%%%%%%%%%%%%%%%%%%%%%%%%%%%%%%%%%
This work was supported in part by NIH grant GM073857 to O.A.A. and Y.K.R.

\appendix

%%%%%%%%%%%%%%%%%%%%%%%%%%%%%%%%%%%%%%%%%%%%%%%
%
\section{Polynomial equations}\label{sec:app-a}
%
%%%%%%%%%%%%%%%%%%%%%%%%%%%%%%%%%%%%%%%%%%%%%%%
The $2\mu$ Eqs.~(\ref{eq:5}) with the $\beta\epsilon_m$ expressed via
parameters (\ref{eq:7}) and (\ref{eq:8}) and the $g_{mm'}$ as stated in
Sec.~\ref{sec:comb} acquire the following form (for $\mu\geq2$):
%\begin{subequations}\label{eq:a1}
\begin{eqnarray}\label{eq:a1a} \fl
  \frac{t}{\tau} &=
  \frac{w_{1}^{2}}{1+w_{1}}\frac{1+w_{2}}{w_{2}}\frac{1+w_{\mu+1}}{w_{\mu+1}},
  \\ \label{eq:a1b} \fl
  t^{2} &=(1+w_{m})\frac{w_{1}^{2}}{(1+w_{1})^{2}}
  \frac{1+w_{m+1}}{w_{m+1}}\frac{1+w_{\mu+m-1}}{w_{\mu+m-1}}
  \frac{1+w_{\mu+m}}{w_{\mu+m}},
  % \nonumber \\ &\hspace{5mm}\times 
  \quad
  m=2,...,\mu-1
  \\ \label{eq:a1c} \fl
  t^{2} &=(1+w_{\mu})\frac{w_{1}^{2}}{(1+w_{1})^{2}}
  \frac{1+w_{2\mu-1}}{w_{2\mu-1}}\frac{1+w_{2\mu}}{w_{2\mu}},
  \\ \label{eq:a1d} \fl
  t &=(1+w_m)\frac{w_{1}}{1+w_{1}},
  \qquad m=\mu+1,...,2\mu.
\end{eqnarray}
%\end{subequations}
From (\ref{eq:a1d}) we infer
\begin{eqnarray}\label{eq:a2} 
 w_{\mu+1}=...=w_{2\mu}\doteq w,
\end{eqnarray}
which, upon substitution, simplifies (\ref{eq:a1a})-(\ref{eq:a1d}) into
%\begin{subequations}\label{eq:a3}
\begin{eqnarray}\label{eq:a3a}
  \frac{t}{\tau} &=
  \frac{w_{1}^{2}}{1+w_{1}}\frac{1+w_{2}}{w_{2}}\frac{1+w}{w},
  \\ \label{eq:a3b}
  t^{2} &=(1+w_{m})\frac{w_{1}^{2}}{(1+w_{1})^{2}}
  \frac{1+w_{m+1}}{w_{m+1}}\frac{(1+w)^{2}}{w^{2}}, 
  % \nonumber \\ &\hspace{35mm} 
  \quad m=2,...,\mu-1,
  \\ \label{eq:a3c}
  t^{2} &=(1+w_{\mu})\frac{w_{1}^{2}}{(1+w_{1})^{2}}
  \frac{(1+w)^{2}}{w^{2}},
  \\ \label{eq:a3d}
  t &=(1+w)\frac{w_{1}}{1+w_{1}}.
\end{eqnarray}
%\end{subequations}
Substitution of (\ref{eq:a3d}) into (\ref{eq:a3a})-(\ref{eq:a3c}) yields 
%\begin{subequations}\label{eq:a4} 
 \begin{eqnarray}
  \label{eq:a4a}
  w_{1} & = \frac{w}{\tau}\frac{w_{2}}{1+w_{2}},
  \quad \tau\neq0,
  \\ \label{eq:a4b}
  w_{m} & = w^{2} \frac{w_{m+1}}{1+w_{m+1}}-1,
  \quad m=2,...,\mu-1, \\ \label{eq:a4c}
  w_{\mu} & = w^{2}-1,
%  \\ \label{}
%  w_{\mu+1} & =...=w_{2\mu} \doteq w.
\end{eqnarray}
%\end{subequations}
which express all $w_m$ for $m\leq\mu$  from $w$ recursively.

Next we show that these recursive relations can be satisfied by Chebyshev
polynomials of the second kind, which themselves are generated recursively
from $S_0(w)=1$ and $S_1(w)=w$ via
\begin{equation}\label{eq:a5} 
S_{m+2}(w)=wS_{m+1}(w)-S_{m}(w),\quad m=0,1,2,\ldots
\end{equation}
We reason inductively by writing (\ref{eq:a4c}) in the form
  \begin{eqnarray}\label{eq:a6} 
    w_{m} %= w^{2}-1 = \frac{S_{2}(w)}{S_{0}(w)} 
    = \frac{S_{\mu-m+2}(w)}{S_{\mu-m}(w)},\quad m=\mu.
  \end{eqnarray}
  If (\ref{eq:a6}) holds for some $m$ then we infer from (\ref{eq:a4b}) that
  it also holds for $m-1$:
 \begin{eqnarray}\label{eq:a7}
    w_{m-1} & = w^{2} \frac{w_{m}}{1+w_{m}}-1
    =  \frac{w^{2}S_{\mu-m+2}(w)}{S_{\mu-m+2}(w)+S_{\mu-m}(w)}-1 \nonumber  \\
    &= \frac{wS_{\mu-m+2}(w)-S_{\mu-m+1}(w)}{S_{\mu-m+1}(w)} 
    = \frac{S_{\mu-m+3}(w)}{S_{\mu-m+1}(w)}.
  \end{eqnarray}
This validates (\ref{eq:a6}) for $m=2,\ldots,\mu$. 
We use (\ref{eq:a4a}) to obtain
\begin{eqnarray} \label{eq:a8}
    w_{1}  = \frac{S_{\mu}(w)}{\tau S_{\mu-1}(w)}.
\end{eqnarray}
The polynomial equation that determines $w$,
  \begin{eqnarray}\label{eq:a9}
    (1+w-t)S_{\mu}(w) = t\tau S_{\mu-1}(w),
  \end{eqnarray}
follows from (\ref{eq:a3d}) substituted in (\ref{eq:a8}).
This completes the derivation of (\ref{eq:9}) and (\ref{eq:10}). 
All $w_m$ must be non-negative to be physically meaningful.
Only one root of (\ref{eq:a9}) satisfies this criterion.

%%%%%%%%%%%%%%%%%%%%%%%%%%%%%%%%%%%%%%%%%%%%%%%
%
\section{Particle population densities}\label{sec:app-b}
%
%%%%%%%%%%%%%%%%%%%%%%%%%%%%%%%%%%%%%%%%%%%%%%%
The solution of the linear Eqs.~(\ref{eq:6}) yields the following explicit
expressions for the population densities of statistically interacting
particles:
\begin{eqnarray}\label{eq:b1} 
  \bar{N}_m 
  & = \frac{S_{\mu-m}(w)S_{\mu-m+1}(w)}{\gamma_\mu}, \nonumber \\ 
  \bar{N}_{\mu+m} 
  & = \frac{[S_{\mu-m}(w)]^2}{\gamma_\mu}, \quad m=1,\ldots,\mu,
\end{eqnarray}
\begin{eqnarray}\label{eq:b2} 
  % \gamma_\mu= &\frac{1+w}{w}\left[2\sum_{m=1}^\mu S_{m-1}(w)S_m(w) \right. \nonumber \\
  % & \hspace*{15mm}\left.+\frac{t-1}{1+w-t}S_{\mu-1}(w)S_\mu(w)\right].
  \gamma_\mu 
  & \doteq (1+w)\!\left[\,\sum_{m=0}^{\mu-1}
    \big[S_m(w)\big]^2+\frac{\big[S_\mu(w)\big]^2}{t\tau}\,\right],
  \nonumber \\
  & =(1+w)\left[\frac{2\mu+1-S_{2\mu}(w)}{4-w^2}
    +\frac{\big[S_\mu(w)\big]^2}{t\tau}\right].
\end{eqnarray}
Entropy (\ref{eq:12}), enthalpy (\ref{eq:13}), helicity (\ref{eq:14}), and
density of segments (\ref{eq:15}) can all be expressed in terms of the
$\bar{N}_m$:
\begin{eqnarray}\label{eq:b3} 
  \frac{\bar{S}}{k_\mathrm{B}}
  =\sum_{m=1}^{2\mu}\bar{N}_m\Big[(1+w_m)\ln(1+w_m)-w_m\ln w_m\Big],
\end{eqnarray}
\begin{equation}\label{eq:b4} 
  \bar{H}=\sum_{m=1}^{2\mu}\bar{N}_m\epsilon_m,
\end{equation}
\begin{equation}\label{eq:b4.1} 
  \bar{N}_\mathrm{seg} = \bar{N}_1
  =\frac{\big[S_\mu(w)\big]^2}{\gamma_\mu}\frac{w+1-t}{t\tau},
\end{equation}
\begin{eqnarray}\label{eq:b5} 
  \bar{N}_\mathrm{hl} 
  =1-\bar{N}_1-2\sum_{m=2}^{\mu}\bar{N}_m 
  -\sum_{m=\mu+1}^{2\mu}\bar{N}_m % \nonumber \\ & 
  = \frac{\big[S_\mu(w)\big]^2}{\gamma_\mu}\frac{w+1}{t\tau},
\end{eqnarray}

The shortest proof of (\ref{eq:b1}) uses its substitution into a scaled
version of (\ref{eq:6}),
\begin{eqnarray}\label{eq:b6} 
  w_{m'}\bar{N}_{m'}
  =\sum_{m=1}^\mu\big(g_{m'm}\bar{N}_m+g_{m',m+\mu}\bar{N}_{m+\mu}\big)
  =\delta_{1m'}.
\end{eqnarray}
We perform this substitution in four batches: (i) for $m'=2\mu$ we use
$g_{2\mu,m}=-\delta_{\mu m}$; (ii) for $m'=\mu+m''$, $m''=1,\ldots,\mu-1$ we
use $g_{\mu+m'',m}=-\delta_{m''m}-\delta_{m'',m-1}$; (iii) for
$m'=2,\ldots,\mu$ we use $g_{m'm}=-\delta_{m',m+1}$; and (iv) for $m'=1$ we
use $g_{1m}=2$, $g_{1,\mu+m}=1$, $m=1,\ldots,\mu$.

In the first three batches (\ref{eq:b6}) is shown to be satisfied by merely
using (\ref{eq:10}) and (\ref{eq:a5}):
% \begin{eqnarray}\label{eq:b7} \fl
%   \mathrm{(i)}:~ \frac{1}{\gamma_\mu}\Big[w[S_0(w)]^2-S_0(w)S_1(w)\Big]=0,
% \end{eqnarray}
% \begin{eqnarray}\label{eq:b8} \fl 
%   \mathrm{(ii)}:~ 
%   &\frac{1}{\gamma_\mu}\Big[w[S_{\mu-m''}(w)]^2 
%   \!-\!S_{\mu-m''}(w)S_{\mu-m''+1}(w) % \nonumber \\ &\hspace{15mm}
%   \!-\!S_{\mu-m''-1}(w)S_{\mu-m''}(w)\Big]=0,
% \end{eqnarray}
% \begin{eqnarray}\label{eq:b9} \fl
%   \mathrm{(iii)}:~ 
%   &\frac{1}{\gamma_\mu}\left[
%     \frac{S_{\mu-m'+2}(w)}{S_{\mu-m'}(w)}
%     S_{\mu-m'}(w)S_{\mu-m+1}(w) % \right. \nonumber \\ &\hspace{20mm}  
%     -S_{\mu-m'+1}(w)S_{\mu-m+2}(w)\right]=0.
% \end{eqnarray}
% In the fourth batch, (\ref{eq:b6}) reduces to the identity,
% \begin{eqnarray}\label{eq:b10}\fl 
%   \mathrm{(iv)}:~ 
%   &\frac{1}{\gamma_\mu} \left[\frac{[S_{\mu}(w)]^2}{\tau}
%     +2\sum_{m=1}^{\mu}S_{m-1}(w)S_m(w) 
%     % \right. \nonumber \\ &\hspace{20mm} \left.
%     +\sum_{m=1}^\mu[S_{m-1}(w)]^2\right]=1,
% \end{eqnarray}
\begin{eqnarray}\label{eq:b7} \fl
  \mathrm{(i)}:~ w[S_0(w)]^2-S_0(w)S_1(w)=0,
\end{eqnarray}
\begin{eqnarray}\label{eq:b8} \fl 
  \mathrm{(ii)}:~ 
  & w[S_{\mu-m''}(w)]^2 
  -S_{\mu-m''}(w)S_{\mu-m''+1}(w) 
  -S_{\mu-m''-1}(w)S_{\mu-m''}(w)=0,
\end{eqnarray}
\begin{eqnarray}\label{eq:b9} \fl
  \mathrm{(iii)}:~ 
  & \frac{S_{\mu-m'+2}(w)}{S_{\mu-m'}(w)}
    S_{\mu-m'}(w)S_{\mu-m+1}(w) 
    -S_{\mu-m'+1}(w)S_{\mu-m+2}(w)=0.
\end{eqnarray}
In the fourth batch, (\ref{eq:b6}) reduces to the identity,
\begin{eqnarray}\label{eq:b10}\fl 
  \mathrm{(iv)}:~ 
  &\frac{[S_{\mu}(w)]^2}{\tau}
    +2\sum_{m=1}^{\mu}S_{m-1}(w)S_m(w) 
    +\sum_{m=1}^\mu[S_{m-1}(w)]^2=\gamma_{\mu},
\end{eqnarray}
which is proven by also using (\ref{eq:9}).

The two sources of disorder identified in Sec.~\ref{sec:heli-ent}, namely the
disorder in the sequence of coil/helix segments of diverse lengths and
disorder within individual coil segments, are related to the population
densities $\bar{N}_m$ of $2\mu$ species of particles from three catgories
(hosts, hybrids, and tags).

Hosts $(m=1)$ generate coil segments out of the helix pseudo-vacuum whereas
hybrids $(m=2,\ldots,\mu)$ and tags $(m=\mu+1,\ldots,2\mu)$ extend coil
segments at the expense of helix segments.  Thermally excited hosts at random
locations along the polypeptide helix thus produce one source of disorder and
germinate the other source of disorder via the thermal excitation of hybrids
and tags nested inside.

Each coil segment, nucleated by exactly one host particle, forms a
self-avoiding random walk assembled from hybrids and tags.  The distribution
of hybrids and tags inside a large coil segment as realized in the limit
$\tau\to0$ at $t<t_0$ and inferred from (\ref{eq:b1}), reads
\begin{eqnarray}\label{eq:b11}
  \bar{N}_{m} 
  &= \frac{2\sin\big((m-1)\phi_0\big)\sin\big(m\phi_0)}{(\mu+1\big)t_{0}}, 
  \nonumber \\
  \bar{N}_{\mu+m} 
  &=\frac{2\sin^{2}\!\big(m\phi_0)}{(\mu+1\big)t_{0}}, 
  \quad m = 1,\ldots,\mu,
\end{eqnarray}
where $\phi_0=\pi/(\mu+1)$. In the limit $\mu\to\infty$, the distributions of
both hybrids and tags acquire identical $\sin^2x$ density profiles if we set
$x=m/\mu$ for hybrids and $x=(\mu+m)/\mu$ for tags.

%\vfill

%\pagebreak

%%%%%%%%%%%%%%%%%%%%%%%%%%%%%%%%%%%%%%%%%%%%%%%
\section*{References}
%\References

\end{document}